\documentclass{amsart}
\pdfoutput=1
\usepackage{amsmath}
\usepackage{graphicx}
\usepackage{enumerate}
\usepackage{natbib}
\usepackage{url}
\usepackage{booktabs}
\usepackage{amsbsy}
\usepackage{amsfonts}
\usepackage{amssymb}
\usepackage{amsmath}
\usepackage{amsaddr}

\newcommand{\bds}[1]{\boldsymbol{#1}}
\newcommand{\bse}{\textbf{\textrm{b}}}
\newcommand{\bsb}{\textbf{\textrm{b}}}
\newcommand{\bsbb}{\textrm{b}}

\newcommand{\code}[1]{\texttt{#1}}
\newcommand{\bsG}{\textbf{\textrm{G}}}
\newcommand{\bsR}{\textbf{\textrm{R}}}

\newcommand{\bsZ}{\textbf{\textrm{Z}}}
\newcommand{\bsW}{\textbf{\textrm{W}}}
\newcommand{\bsX}{\textbf{\textrm{X}}}

\newcommand{\bsA}{\boldsymbol{A}}

\newcommand{\bsz}{\textbf{\textrm{z}}}
\newcommand{\bsx}{\textbf{\textrm{x}}}

\newcommand{\bsy}{\textbf{\textrm{y}}}
\newcommand{\bsv}{\textbf{\textrm{v}}}
\newcommand{\bsc}{\boldsymbol{c}}

\newcommand{\tr}{\textrm{tr}}
\newcommand{\var}{\textrm{var}}

\newcommand{\moparm}{\boldsymbol{\Psi}}
\newcommand{\bsSigma}{\boldsymbol{\Sigma}}
\newcommand{\bsbeta}{\boldsymbol{\beta}}
\newcommand{\bsGamma}{\boldsymbol{\Gamma}}
\newcommand{\te}[1]{\textrm{#1}}
\newcommand{\der}[1]{\frac{\partial}{\partial #1}}

\newcommand{\dert}[2]{\frac{\partial #2}{\partial #1}}
\newcommand{\dertwot}[2]{\frac{\partial^2 #2}{\partial #1^2}}
\newcommand{\derthreet}[2]{\frac{\partial^3 #2}{\partial #1^3}}
\newcommand{\derfourt}[2]{\frac{\partial^4 #2}{\partial #1^4}}
\newcommand{\be}[1]{\boldsymbol{e}_{#1}}
\newcommand{\bep}[1]{\boldsymbol{e}_{#1}^{\prime}}
\newcommand{\ud}{\mathrm{d}}

\usepackage{graphicx}

\usepackage{amssymb}
\begin{document}

\author[Karl, Yang, and Lohr]{Andrew T. Karl}
\address{Adsurgo LLC, Denver, CO}
\author{Yan Yang}
\address{Arizona State University, Tempe, AZ}
\author{Sharon L. Lohr}
\address{Westat, Rockville, MD}

\title[MLE for Non-nested, Multiresponse GLMMs]{Computation of Maximum Likelihood Estimates for Multiresponse Generalized Linear Mixed Models with Non-nested, Correlated Random Effects}

\begin{abstract}
Estimation of generalized linear mixed models (GLMMs) with non-nested random effects structures requires approximation of high-dimensional integrals. Many existing methods are tailored to the low-dimensional integrals produced by nested designs. We explore the modifications that are required in order to adapt an EM algorithm with first-order and fully exponential Laplace approximations to a non-nested, multiple response model. The equations in the estimation routine are expressed as functions of the first four derivatives of the conditional log likelihood of an arbitrary GLMM, providing a template for future applications. We apply the method to a joint Poisson-binary model for ranking sporting teams, and discuss the estimation of a correlated random effects model designed to evaluate the sensitivity of value-added models for teacher evaluation to assumptions about the missing data process. Source code in R is provided in the online supplementary material.
\end{abstract}

\maketitle

\section*{Revision note (22 October 2026)}
This arXiv update records corrections found by the first author, Andrew T. Karl, while checking his published work with Codex GPT-6 Astra. The changes to the paper are relatively minor and leave the overall method and most of the text unchanged. For this reason, the first author chose to update the arXiv paper rather than submit a correction to the journal.

The main formula correction is to the Hessian in Equation~(\ref{eq:sigma}). For the response models treated here, the omitted mixed-derivative terms are needed only when normal responses have correlated residual errors. The applications and simulations use conditionally independent observations, so this omission does not affect their Hessians. The normal-response implementations inspected in \code{mvglmmRank}, \code{GPvam}, and \code{RealVAMS} already use the correct matrix calculation, so the printed error did not carry through to that software. The normal log likelihood is quadratic in the linear predictor, so its third- and fourth-order derivatives are zero. Appendices A and B now give the calculations in matrix form.

Other edits fix notation, the perturbed-mode definition, the Newton stopping test, and small numerical errors, and clarify scores and convergence. Separate bugs in the supplemental programs have also been corrected and documented in their README. Limited paired checks changed some first-order estimates and fitted neutral-site win probabilities. Their effect on the published tables is still unclear: the supplied football data and example model differ from those described in the paper, and the archived supplement lacks the original simulation records. The tables are retained as published.

\section*{NOTICE}
This is the author's version of a work that was accepted for publication in \textit{Computational Statistics \& Data Analysis}. Changes resulting from the publishing process, such as peer review, editing, corrections, structural formatting, and other quality control mechanisms may not be reflected in this document. Changes may have been made to this work since it was submitted for publication. A definitive version was subsequently published in \textit{Computational Statistics \& Data Analysis}, 73 (May 2014), 146--162, DOI:10.1016/j.csda.2013.11.019


\section{Introduction}
Generalized linear mixed models (GLMMs) are popular for fitting data sets with correlated responses from a member of the exponential family, where random effects are used to account for within-subject correlation. They have been used in applications ranging from ecology and evolution \citep{bolker} to epidemiology \citep{brumback} to education \citep{broatch10}.
Computing maximum likelihood (ML) estimates in GLMMs is more complicated than comparable computations in linear mixed models because the log-likelihood function for the parameters in a GLMM contains integrals over the random effects that must be computed numerically. The dimension of the integration depends on the structure of the random effects.
Many of the computational methods in current use for GLMMs concentrate on nested generalized linear mixed models, in which the units at the top level of the hierarchy are independent. The nested structure considerably simplifies the computation of ML estimates because it reduces the dimensions of the integrals that need to be calculated and simplifies the inversion of relevant covariance matrices.

Many longitudinal studies of subjects in social sciences or epidemiology, however, involve a non-nested structure of the random effects. In education, students proceed through different classrooms in their academic career. Thus, the response of a student at time $t$ may depend on the full set of classes the student has taken up to that time.
This leads to a multiple membership structure \citep{browne01} in which each student is associated with multiple random classroom effects in sequence.
The correlation structure becomes complex in multiple membership models because one student's response is correlated with responses of all the students with whom the student shares classes. Each student may have multiple and mixed responses at each time: for example one response might be a score on a standardized test, while another response might be whether the student is required to attend summer school. Furthermore, allowing binary responses allows for the modeling of a missing data mechanism: one of the responses is whether the student is present for the exam; if the student is absent, then his test score is missing. Jointly modeling the student test scores and missing data mechanism allows for a sensitivity analysis to assumptions about the dropout process \citep{karlcpm}.

Multiple membership models have also been used in longitudinal studies of health outcomes. \citet{chandola} use them to examine the effects of neighborhood and family on health outcomes, when individuals can move from one household to another and from one neighborhood to another over time. \citet{gue} discuss the advantages of jointly modeling continuous and ordinal measures of depression over time. Accounting for patients seeing different therapists over time would lead to a non-nested structure for the random effects. In a different type of application, \citet{page} study the relationship between the home field advantage and effects of different referees in soccer.  The home field advantage of a team for a specific referee is modeled by a random effect, and it would be desirable to also include the home and away teams as random effects. Typically, the number of goals scored in soccer games is low, and a Poisson distribution may be better for fitting the response than a normal distribution.

All of these applications share the following structure. At each time period, there is a potentially multivariate outcome for each subject where at least one of the responses has a non-normal distribution. Each subject is associated with multiple higher-level units, or each observation is a function of multiple levels of the same random factor: a student belongs to multiple teachers or classrooms, a person belongs to different neighborhoods over time, a patient sees different therapists, a referee officiates for different teams and at different locations, and game outcomes depend on combinations of different teams' effects. The multiple membership pattern induces a complex correlation structure in which each observation of a subject is potentially correlated with those of every other subject sharing any of the higher-level units.  In a GLMM, the dimensionality of the integration is related to the complexity of the random effects design matrix. When this matrix cannot be factored across subjects, the integral in the log-likelihood will have the same dimension as the number of random effects in the model. In addition, the random effects associated with different responses for the same subject are correlated. The parameters of interest are the fixed regression coefficients, the empirical best linear unbiased predictors (EBLUPs), the variance components for low- and high-level units, and the correlations among the variance components.
\citet{pinheiro06} note the need for an efficient estimation routine for non-nested models such as these.

In this paper, we propose a general method based on the Expectation-Maximization (EM) algorithm for computing ML estimates in multiresponse GLMMs with correlated, potentially non-nested random effects. We exploit the sparse structure of the random effects design matrices while using first-order and fully exponential Laplace approximations for the multidimensional integrals to furnish a balance of accuracy and speed, providing an improvement over the widely used pseudo-likelihood model approximation. Furthermore, we express the estimation routine in terms of the first four derivatives of the conditional log likelihood for each response, facilitating its application to new problems. Estimation routines for GLMMs tend to be more application-dependent than routines available for linear mixed models; this paper presents a template for jointly estimating multiple GLMMs without assuming any special structure for the random effects design matrices corresponding to each response.
 R code to fit a non-nested Poisson-binary model and a multiple membership binary response is presented in the online supplement, along with sample data. The code provides a foundation for readers wishing to apply the routine in new settings, since the challenging programming aspects have already been addressed.

At the time of this revision, maintained R implementations are available in \code{mvglmmRank} (\url{https://CRAN.R-project.org/package=mvglmmRank}) for sports ranking models and in \code{glmmFEL} (\url{https://CRAN.R-project.org/package=glmmFEL}) for binary logit/probit and Poisson models specified by design matrices with independent random effects sharing a single variance. The latter implements Laplace approximations and fully exponential corrections to conditional means and the diagonal entries of the conditional covariance matrix; see the package manuals for the supported models and options. For a subsequent application of this framework to joint sporting outcomes, see \citet{broatchkarl17}, who develop normal--binary and Poisson--binary models and describe their implementation in \code{mvglmmRank}.

In Section~\ref{sec:cre}, we describe the multivariate generalized linear mixed model treated in this paper, illustrate the model with a Poisson-binary example, and summarize previous work on computational methods. Section~\ref{sec:estimation} presents the estimation of the joint model via an EM algorithm. Section~\ref{sec:computation} discusses the techniques
that are necessary in order to compute the EM estimates for large data sets.  Applications to sports rankings and student test scores are given in Section~\ref{sec:application}. A simulation study to evaluate the performance of various approximations appears in Section~\ref{sec:sim}. Section~\ref{sec:summary} discusses the results and outlines possible extensions.

\section{Multiresponse Generalized Linear Mixed Model}\label{sec:cre}
In this section we describe the model adopted for the potentially non-nested models studied in this paper. We adopt a correlated random effects model \citep{fieuws06,lin09,karlcpm} in which the multiple observations on a subject are assumed to be independent after conditioning on distinct, but correlated, random effects. This is an extension of a shared parameter model \citep{wu88,riz09} in which the different responses of a subject share the same random effects.

\subsection{The Joint Model}
\label{sec:jointmodel}
Suppose there are $n$ subjects, and between 1 and $q$ response variables are measured on each subject. The data may be unbalanced: it is possible that subjects may have repeated or missing measurements on any one of the $q$ response variables. Let $n_{(i)}$ be the number of observations on response variable $i$, and let $\bsy_{(i)}$ denote the vector of those measurements. Response $i$ may be either continuous or discrete, and different responses may be of different types. If only one response is measured, $q=1$ and the vector $\bsy_{(1)}$ consists of the data for all subjects.
Let $\bse_{(1)},...,\bse_{(q)}$ be distinct yet possibly correlated vectors of random effects which are concatenated (and possibly permuted) into a single vector, $\bse$, and assume that $\bse  \sim N(\bds{0},\bsG)$. The random effects $\bse_{(i)}$ may be used to model (potentially different) multi-level structures in the response variables. We do not assume any particular structure for $\bsG$, though $\bse$ may typically be organized so that $\bsG$ is block diagonal, as discussed in Section~\ref{sec:mstep}.
Also assume that after conditioning on the random effect vectors, $\bsy_{(1)}|\bse_{(1)},...,\bsy_{(q)}|\bse_{(q)}$ are independent.
Let $f(\bsy_{(i)}|\bse_{(i)})$ represent the conditional density function for the $i$-th response variable, determined by the distribution and link function used for that response variable. The joint likelihood may thus be expressed as
\begin{align}
f(\bsy_{(1)},\ldots,\bsy_{(q)})&=\int{f(\bsy_{(1)}|\bse_{(1)})\ldots f(\bsy_{(q)}|\bse_{(q)})f(\bse)\ud \bse}\nonumber\\
&=\int{f(\bsy_{(1)}|\bse)\ldots f(\bsy_{(q)}|\bse)f(\bse)\ud \bse}\label{eq:cpmlik}.
\end{align}

The $l$-th observation for response variable $i$, denoted $y_{l(i)}$, has expected value $\mu_{l(i)}$, where for some link function $g$, $\eta_{l(i)} = g(\mu_{l(i)}) = \bsx_{l(i)}^{\prime} \bsbeta_{(i)} + \bsz_{l(i)}^{\prime} \bse$. Thus $\bsx_{l(i)}^{\prime}$ represents the $l$-th row of the design matrix $\bsX_{(i)}$ for response variable $i$, and $\bsbeta_{(i)}$ is the vector of fixed effects for response variable $i$. The vector $\bsz_{l(i)}$ has length equal to that of $\bse$, and contains a zero corresponding to elements of $\bse$ other than $\bse_{(i)}$. This definition of $\bsz_{l(i)}$ streamlines our discussion of the estimation routine. We assume the matrices $\bsX_{(i)}$ are full rank, but we do not assume any special structure for the matrix $\bsZ_{(i)}$ which is formed with rows $\bsz_{l(i)}^{\prime}$. The matrices $\bsZ_{(i)}$ will typically be sparse, and we capitalize on this sparseness in programming the algorithm to make it applicable to large data sets, as discussed in Section~\ref{sec:computation}.

For a normally distributed response $\bsy_{(i)}$, the error terms are modeled in a covariance matrix $\bsR_{(i)}=cov(\bsy_{(i)}|\bse)$.  This matrix may range in structure from a multiple of an identity matrix to an unstructured matrix. However, for single parameter distributions from the exponential family the variance is a function of the mean and not explicitly modeled in a GLMM \citep{stroup}.

\subsection{Example: Poisson and Binary Responses}\label{ssec:ME}
The joint model in Section~\ref{sec:jointmodel} can accommodate a wide variety of situations. In order to make some of the concepts more concrete, we now provide an example using sports teams that we will follow throughout the paper.
Suppose that there are $p$ teams that play in $n$ games. Two types of responses are recorded for each game $g$, $g=1,\ldots,n$: team scores $\bsy_{(1)}$, and binary win/loss indicators $\bsy_{(2)}$. We consider sports for which ties are necessarily settled in overtime. Typically, teams are ranked using either the scores or the win-loss records \citep{football}; the multiresponse model (\ref{eq:cpmlik}) allows rankings to be done using both.
\citet{annis} consider a similar approach for ranking college football teams.

We model the $2n$ team scores contained in $\bsy_{(1)}$ using a Poisson distribution. We assign 1 to the variable $y_{g(2)}$  with a ``home team'' win in game $g$ and  0 otherwise, producing a vector $\bsy_{(2)}=(y_{1(2)},\ldots,y_{n(2)})^{\prime}$. For neutral site games, the ``home team'' may be chosen arbitrarily: in this example, we only use this distinction to keep track of the win/loss modeled in $y_{g(2)}$.

In a game between teams A and B, the probability of a victory for team A is modeled as a function of the difference between ``win propensity ratings'' for A and B. Team A's score is modeled as a function of the difference between team A's ``offensive rating'' and team B's ``defensive rating.'' Team B's score is modeled similarly. The offensive, defensive, and win propensity ratings of the $j$-th team for $j=1,\ldots,p$ are assumed to be random effects $\bsbb_j^o$, $\bsbb_j^d$, and $\bsbb_j^w$ with $\bsb_j=(\bsbb_j^o,\bsbb_j^d,\bsbb_j^w)^{\prime}\sim N_3(\bds{0},\bsG^*)$, where $\bsG^*$ is an unstructured covariance matrix and $p$ represents the number of teams being ranked.  In addition, $\bsb\sim N(\bds{0},\bsG)$, where $\bsb=(\bsb_1^{\prime},\ldots,\bsb_p^{\prime})^{\prime}$ and $\bsG$ is block diagonal with $p$ copies of $\bsG^*$. To match the notation of Section~\ref{sec:cre}, $\bsb_{(1)}=(\bsbb_1^o, \bsbb_1^d,\bsbb_2^o, \bsbb_2^d,\ldots,\bsbb_p^o, \bsbb_p^d)^{\prime}$, $\bsb_{(2)}=(\bsbb_1^w, \bsbb_2^w, \ldots,\bsbb_p^w)^{\prime}$, $n_{(1)}=2n$, and $n_{(2)}=n$. Notice how the components of $\bsb_{(2)}$ are folded into those of $\bsb_{(1)}$ when forming $\bsb$ in order to ensure that $\bsG$ is block diagonal.

Assume the conditional team scores $\bsy_{(1)}|\bse=\bsy_{(1)}|\bse_{(1)}$ follow a Poisson distribution, and that the conditional outcomes $\bsy_{(2)}|\bse=\bsy_{(2)}|\bse_{(2)}$ follow a Bernoulli distribution. The fixed effect $\beta_{(1)}$ for the score response models the mean score, and $\bsX_{(1)}$ is a column of 1's. We consider a parsimonious model with a single mean score. Richer models could explore differences in mean home team and away team scores. As such, no fixed effects are modeled for $\bsy_{(2)}$.

The random effects design matrix $\bsZ^*_{(1) }$ is a $2n \times 2p$ matrix that tracks which team's offense was matched against which team's defense to produce each of the $2n$ scores. If $y_{l(1)}$ is the score compiled by the offense of team A against the defense of team B, then $\bsz^*_{l(1)}$ contains  1 in the position corresponding to the position of the offensive effect of team A, $\bsbb^o_A$, in $\bsb_{(1)}$, and  $-1$ in the position corresponding to the position of the defensive effect $\bsbb^d_B$ of team B, producing a partially crossed design. $\bsZ^*_{(1)}$ is then expanded into a $2n \times 3p$ matrix $\bsZ_{(1)}$ by inserting columns of zeros into every third column, corresponding to the positions of $\bsb^w$ in $\bsb$. Likewise, $\bsZ^*_{(2) }$ is an $n \times p$ matrix that indicates which teams played each other in each of the $n$ games. Each row  $\bsz^{*\prime}_{g(2)}$ contains  1 in the column corresponding to the ``home team'' and  $-1$ in the column corresponding to the ``away team''. $\bsZ^*_{(2)}$ is expanded into an $n \times 3p$ matrix $\bsZ_{(2)}$ by inserting columns of zeros in positions corresponding to the locations of $\bsb^o$ and $\bsb^d$ in $\bsb$.

To illustrate the construction of these matrices, consider a simple example with two teams: A and B. Suppose team A wins a home game against team B by a score of 17 to 10 and then loses a game while visiting team B by a score of 21 to 24. Then $\bsy_{(1)}=\left(17,10,21,24\right)^{\prime}$ and $\bsy_{(2)}=\left(1,1\right)^{\prime}$. Again letting $o,d,w$ represent the offensive, defensive, and win propensity effects, the random effects structure produced by this example is $\bse=\left(\bsbb_A^o,\bsbb_A^d,\bsbb^w_A,\bsbb_B^o,\bsbb_B^d,\bsbb^w_B\right)^{\prime}$,
\begin{align*}
\bsG^*=&\begin{pmatrix}var(o)&cov(o,d)&cov(o,w)\\cov(d,o)&var(d)&cov(d,w)\\cov(w,o)&cov(w,d)&var(w)\end{pmatrix},\\
\bsG=&\begin{pmatrix}\bsG^*&\bds{0}\\ \bds{0}&\bsG^*  \end{pmatrix},\\
\bsZ_{(1)}=&\begin{pmatrix}1&0&0&0&-1&0\\0&-1&0&1&0&0\\1&0&0&0&-1&0\\0&-1&0&1&0&0 \end{pmatrix},\\
\bsZ_{(2)}=&\begin{pmatrix}0&0&1&0&0&-1\\0&0&-1&0&0&1 \end{pmatrix}.
\end{align*}

The likelihood functions for the scores (using a log link) and outcomes (using a probit link, with $\Phi$ representing the standard normal cumulative distribution function) are
\begin{align}
f(\bsy_{(1)}|\bse)=&\prod_{l=1}^{2n} \left[\frac{1}{y_{l(1)}!}\te{ exp}\left\{y_{l(1)}(\beta_{(1)}+\bsz_{l(1)}^{\prime}\bse)\right\}\te{exp}\left\{-\te{exp}\left[\beta_{(1)}+\bsz_{l(1)}^{\prime}\bse\right]\right\}\right]\label{eq:Poisson}\\
f(\bsy_{(2)}|\bse)=&\prod_{g=1}^n \left[\Phi\left\{\left(-1\right)^{1-y_{g(2)}}\bsz_{g(2)}^{\prime}\bsb\right\}\right]\label{eq:binary}.
\end{align}
Typically, teams would be ranked by maximizing only one of the likelihoods, (\ref{eq:Poisson}) or (\ref{eq:binary}). The joint likelihood function combines both to produce
\begin{align}\label{eq:joint}
L(\beta_{(1)},\bsG)&=\idotsint f(\bsy_{(1)}|\bse) f(\bsy_{(2)}|\bse) f(\bse) \ud \bse,
\end{align}
where $f(\bse)$ is the density of $\bse\sim N(\bds{0},\bsG)$. The centerpiece of the joint model is the pair of unique off-diagonal covariance terms between $\bsb_{(1)}$ and $\bsb_{(2)}$ in the $\bsG$ matrix, referred to as $cov(w,o)$ and $cov(w,d)$ in the example above. To illustrate: if offensive ratings are highly correlated with the win propensity ratings, then the teams with the best offenses would have larger win propensity ratings in the joint model than they would if only (\ref{eq:binary}) were maximized. If these covariance terms were constrained to 0, the resulting model fit would be equivalent to the one obtained by modeling the two responses independently.

\subsection{Computational Methods for GLMMs}
The likelihood in (\ref{eq:cpmlik}) contains an integral that potentially has the same dimension as the number of random effects.
While the marginal likelihood function in completely hierarchical models may be factored into a product of low dimensional integrals and evaluated by quadrature methods \citep{evans95},  models with a potentially non-nested structure (e.g.~multimembership or cross-classified) produce likelihood functions that must be expressed as a single, high-dimensional, intractable integral. Quadrature methods are infeasible for high-dimensional integrals.

Markov Chain Monte Carlo (MCMC) methods are frequently used in a Bayesian approach for these problems. These, however, often require a subjective prior distribution, and the estimates for random effects models can be very sensitive to the choice of prior distribution, as in \citet{mariano10}. Judging convergence for these models can be challenging \citep[p.~357]{handbook}. A method for computing ML estimates avoids the need for a prior distribution, and provides a useful check on sensitivity to the prior even when a Bayesian approach is preferred.

Asymptotic methods such as pseudo-likelihood (PL) linearization \citep{wolfinger93}, penalized quasi-likelihood (PQL, \citeauthor{breslow93}	 \citeyear{breslow93}), first-order Laplace approximations, and fully exponential Laplace approximations \citep{tierney86, tierney89} avoid the problems of MCMC methods and the need for prior distributions, though potentially at the cost of accuracy. \citet{pinheiro95} compare PQL, the first-order Laplace approximation, Gaussian quadrature, adaptive Gaussian quadrature (AGQ), and importance sampling, and conclude that the Laplace approximation and AGQ approximations give the ``best mix of efficiency and accuracy.'' \citet{raud} propose an alternative method for nested GLMMs that uses a sixth-order Laplace approximation.  \citet{pinheiro06} provide scalable Laplacian and AGQ methods for multilevel models with an arbitrary number of nested effects, noting the need for extension to non-nested models.

Most of the work described above uses Newton-Raphson (NR) to find the parameter values maximizing the likelihood. Alternatively, an EM algorithm may be used to estimate the model parameters \citep{dempster77,embook}. EM algorithms treat the random effects as missing data: the E step calculates the conditional expectation of the complete-data log-likelihood, given the observed data and current parameter estimates, and the M step maximizes the conditional expectation of the complete data log-likelihood \citep{laird82}. Although \citet{lindstrom88} highlight advantages of NR methods over the EM algorithm, including quadratic as opposed to linear convergence rates, the EM algorithm provides a viable solution in situations where NR algorithms may fail.  For some applications of correlated random effects models, strong correlations in the random effects covariance matrix cause difficulty for Newton methods \citep{karlem}. The EM algorithm provides a stable maximization routine even when presented with nearly perfectly correlated effects.

When estimating nonlinear mixed models with an EM algorithm, the intractable integral in the marginal likelihood produces a similarly intractable integral in the E step. \citet{steele} uses a fully exponential Laplace approximation in the E step of an EM algorithm to estimate GLMMs with uncorrelated random effects,
and \citet{riz09} apply this approach to a nested design and relax the assumption of a diagonal random effects covariance matrix. Their estimation routine is available via the R \citep{R} package JM \citep{rizjs}.

We use the EM algorithm to compute maximum likelihood estimates of the parameters, using fully exponential and first-order Laplace approximations for the integrals in the E step. The innovations in this approach are (1) a uniform method for the E step that depends on the derivatives of the conditional log likelihood function, which allows the method to be applied to other GLMM settings merely by changing those derivatives, (2) using the sparseness of the random effects design matrices to keep computations manageable, and (3) describing how redundant calculations may be identified and omitted in the presence of non-nested random effects.

\section{Estimation Procedure}\label{sec:estimation}

 Let $\moparm$ be a vector of the unique model parameters. We will refer to $f(\bsy_{(1)},\ldots,\bsy_{(q)}; \moparm)$ as the observed data density function and $f(\bsy_{(1)},\ldots,\bsy_{(q)}, \bse; \moparm) = f(\bsy_{(1)}|\bse; \moparm)\cdots f(\bsy_{(q)}|\bse; \moparm) f(\bse; \moparm)$ as the complete data density function.
Given initial values for the parameters and the random effects, the EM algorithm \citep{dempster77,laird82} alternates between an expectation (E) step and a maximization (M) step. At iteration $(k + 1)$, the E step calculates the conditional expectation of the complete data log-likelihood, given the observed data, ($\bsy_{(1)},\ldots,\bsy_{(q)})$, and parameter estimates obtained in the $k$-th step, $\moparm^{(k)}$. That is, the E step computes
\begin{equation}\label{eq:em2}
Q(\moparm; \moparm^{(k)}) = \int{\left\{\left[\sum_{i=1}^q\log f\left(\bsy_{(i)}|\bse; \moparm\right)\right] + \log f\left(\bse; \moparm\right)\right\} f(\bse|\bsy_{(1)},\ldots,\bsy_{(q)}; \moparm^{(k)})} \ud\bse.
\end{equation}
The M step then maximizes $Q(\moparm; \moparm^{(k)})$ with respect to $\moparm$, resulting in the updated parameter vector $\moparm^{(k + 1)}$ satisfying
\begin{equation}
\int \frac{\partial}{\partial\moparm} \left\{\left[\sum_{i=1}^q\log f\left(\bsy_{(i)}|\bse; \moparm\right)\right]  + \log f(\bse; \moparm)\right\} f(\bse|\bsy_{(1)},\ldots,\bsy_{(q)}; \moparm^{(k)} )\ud\bse\Big|_{\moparm=\moparm^{(k+1)}} = \bds{0}\label{eq:emscore1}.
\end{equation}
Under regularity conditions permitting differentiation under the integral \citep{lehmann}, the gradient of $Q(\moparm;\moparm^{(k)})$ equals the observed data score $S(\moparm)=\left(\partial/\partial\moparm\right)l(\moparm;\bsy_{(1)},\ldots,\bsy_{(q)})$ when both parameter arguments are equal \citep{louis,embook}:
\[
\left.\frac{\partial Q(\moparm;\moparm^{(k)})}{\partial\moparm}\right|_{\moparm=\moparm^{(k)}}=S(\moparm^{(k)}).
\]
The M step in (\ref{eq:emscore1}) instead holds the conditioning parameter fixed at $\moparm^{(k)}$ while optimizing over $\moparm$. Thus an interior fixed point of the exact EM algorithm is a stationary point of the observed data likelihood; an intermediate M step need not set the observed data score to zero. An EM algorithm is used to develop an efficient routine for calculating maximum likelihood estimates of multiple membership linear mixed models (LMMs) under an assumption of missing at random (MAR) by \citet{gpvam,karlem}. When at least one response is non-normally distributed, the integral in $Q(\moparm; \moparm^{(k)})$ has no closed form solution. Because we consider potentially non-nested designs, the dimension of the integral is equal to the length of $\bse$. For GLMMs, \citet{steele} proposes approximating the intractable integrals in (\ref{eq:em2}) with a fully exponential Laplace approximation \citep{tierney89}. \citet{riz09} use this method to fit a nested shared parameter model for a normal and a time-to-dropout process. To estimate the parameters of the correlated random effects model (\ref{eq:cpmlik}), we generalize the approach for non-nested LMMs proposed by \citet{karlem} by incorporating elements of the E step approximation used by \citet{riz09}. The fully exponential Laplace approximation produces computationally expensive corrections to the first-order Laplace approximation. These corrections are not particularly burdensome for the nested design of \citet{riz09} since, in that application, only around 15 thousand such corrections need to be calculated during each iteration. Our aim in this paper is to present the calculation of the model terms for a non-nested design and to discuss computational methods that are required in situations where a naive approach would lead to the calculation of millions of corrections per iteration.

\subsection{Derivatives of the Conditional Log-Likelihood(s)}\label{ssec:derivatives}
Let $\eta_{l(i)}=\bsx^{\prime}_{l(i)}\bsbeta_{(i)}+\bsz^{\prime}_{l(i)}\bse$ be the linear predictor (from the GLMM) for the $l$-th observation for response $\bsy_{(i)}$.
The EM algorithm requires derivatives of $\log f(\bsy_{(i)}|\bse)$ with respect to $\bse$. We use the fact that $\eta_{l(i)}$ is linear in $\bse$ to simplify this calculation. By the chain rule,
\begin{equation}\label{eq:chain}
\frac{\partial \log f(\bsy_{(i)}|\bse)}{\partial\bse}=\sum_{l=1}^{n_{(i)}}\frac{\partial \log f(\bsy_{(i)}|\bse)}{\partial\eta_{l(i)}}\cdot\bsz_{l(i)}
\end{equation}
A total of four derivatives of $\log f(\bsy_{(i)}|\bse)$ with respect to $\eta_{l(i)}$ will be required. Appendix A lists these derivatives for Poisson, binary, and normal distributions.

\subsection{The M step}\label{sec:mstep}
The M step of the EM algorithm maximizes the conditional expectation of the complete data log-likelihood. Let $\widetilde{\bse} = \te{E}[\bse|\bsy_{(1)},\ldots,\bsy_{(q)};\moparm]$ and $\widetilde{\bsv} = \te{var}[\bse|\bsy_{(1)},\ldots,\bsy_{(q)};\moparm]$
represent the conditional expectation and variance, respectively, of $\bse$. These quantities are calculated in the E step and remain fixed during the M step.

\subsubsection{Fixed Effects}
Assuming the fixed effects for each response are distinct, write the M step score for $\bsbeta_{(i)}$ as follows, with the conditional density held fixed at the preceding E step parameter values:
\begin{align}
S(\bsbeta_{(i)})=&\int \dert{\bsbeta_{(i)}}{\log[f(\bsy_{(i)}|\bse)]}f(\bse|\bsy_{(1)},\ldots,\bsy_{(q)})\ud \bse \nonumber\\
=&\sum_{l=1}^{n_{(i)}}\bsx_{l(i)}\int{\dert{\eta_{l(i)}}{\log[f(\bsy_{(i)}|\bse)]}}f(\bse|\bsy_{(1)},\ldots,\bsy_{(q)})\ud \bse \label{eq:scmumiss}
\end{align}
When $\bsy_{(i)}$ is normally distributed with error covariance matrix $\bsR_{(i)}$, the solution for $S(\widehat{\bsbeta}_{(i)})=0$ is
\begin{align}
\widehat{\bsbeta}_{(i)}&=\left(\bsX_{(i)}^{\prime}\bsR_{(i)}^{-1}\bsX_{(i)}\right)^{-1}\bsX_{(i)}^{\prime}\bsR_{(i)}^{-1}\left(\bsy_{(i)}-\bsZ_{(i)}\widetilde{\bse}\right) . \nonumber
\end{align}
By contrast, there is no such closed form solution when $\bsy_{(i)}$ is non-normally distributed since the integrals in (\ref{eq:scmumiss}) are intractable. The approximations to the integrals in (\ref{eq:scmumiss}) will be obtained in the E step. $\widehat{\bsbeta}_{(i)}$ may then be found by Newton-Raphson: the Hessian $\mathcal{H}(\widehat{\bsbeta}_{(i)})$ is calculated by applying a central difference approximation to $S(\bsbeta_{(i)})$ at $\widehat{\bsbeta}_{(i)}$. Iteration $p+1$ of the approximation yields
\[\widehat{\bsbeta}_{(i)}^{p+1} = \widehat{\bsbeta}_{(i)}^{p} - \mathcal{H}(\widehat{\bsbeta}^p_{(i)})^{-1}S(\widehat{\bsbeta}_{(i)}^p).\]
With a recommendation of $\alpha=10^{-8}$, and a negative-definite Hessian near the maximizing solution, iterations continue until
\begin{equation}
-S(\widehat{\bsbeta}_{(i)}^p)^{\prime} \mathcal{H}(\widehat{\bsbeta}^p_{(i)})^{-1}S(\widehat{\bsbeta}_{(i)}^p) < \alpha.\label{eq:nrconv}
\end{equation}

\subsubsection{The Random Effects Covariance Matrix}\label{sssec:gmat}
The M step update for $\bsG$ does not depend on the response distributions included in (\ref{eq:cpmlik}). Throughout Section~\ref{sssec:gmat}, assume that $U$ independent groups of random effects (e.g. levels in a multilevel model) are modeled across combinations of the $q$ response variables. Within each group $u$, $u=1,\ldots,U$, $V_u$ variance components are modeled with $M_u$ random effects per variance component. These effects are correlated within each group.  In the context of the Poisson-binary model in Section~\ref{ssec:ME}, $U=1$, $q=2$, $V_1=3$, and $M_1=p$. An additional game-level effect could be added to the Poisson model for scores to account for intra-game correlation, in which case $U=2$, $V_2=1$, and $M_2=n$.

Let $\bse^{(u,1)},\ldots,\bse^{(u,M_u)}$ be the length-$V_u$ sub-vectors of $\bse$ representing the effects in group $u=1,\ldots,U$. Then $\bse$ may be arranged so that
\begin{align*}
\bse=&\left((\bse^{(1,1)})^{\prime},\ldots,(\bse^{(1,M_1)})^{\prime},(\bse^{(2,1)})^{\prime},\ldots,(\bse^{(2,M_2)})^{\prime},\ldots,(\bse^{(U,1)})^{\prime},\ldots,(\bse^{(U,M_U)})^{\prime}\right)^{\prime}\\
\bsG=&\te{block diagonal}\left(\bsGamma_1,\ldots,\bsGamma_1,\ldots,\bsGamma_2,\ldots,\bsGamma_2,\ldots,\bsGamma_U,\ldots,\bsGamma_U\right).
\end{align*}
$\bsG$ contains $M_u$ copies each of the $V_u\times V_u$ matrix $\bsGamma_u=cov(\bse^{(u,j)})$, for $j=1,\ldots,M_u$. Let the components of $\widetilde{\bse}$ corresponding to $\te{E}[\bse^{(u,j)}|\bsy_{(1)},\ldots,\bsy_{(q)};\moparm]$ be denoted $\widetilde{\bse}^{(u,j)}$ for $u=1,\ldots,U$ and $j=1,\ldots,M_u$. Likewise, let $\widetilde{\bsv}^{(u,j)}$ represent the block of the matrix $\widetilde{\bsv}$ corresponding to $\te{var}[\bse^{(u,j)}|\bsy_{(1)},\ldots,\bsy_{(q)};\moparm]$. When the $\bsGamma_u$ are unstructured, a generalization of steps illustrated by \citet{karlem} shows
\begin{equation}
\widehat{\bsGamma}_{u}=\frac{1}{M_u}\sum_{j=1}^{M_u}\left\{\widetilde{\bsv}^{(u,j)}+\widetilde{\bse}^{(u,j)}\left(\widetilde{\bse}^{(u,j)}\right)^{\prime}\right\}\label{eq:gamM}
\end{equation}
Alternative structures may be imposed on the $\bsGamma_u$ using the same approach as that demonstrated for an AR(1) (first-order autoregressive) $\bsR$ matrix in the next section. This is a general approach: all that is required is the first derivative of $\bsG$ with respect to each unique parameter in the matrix. This also allows for the possibility of a non-block-diagonal $\bsG$.  For the E step, the groupings of random effects (e.g. levels of a multilevel model) do not need to be tracked explicitly, since this information is contained in the structure of $\bsG$ and components of the $\bsZ_{(i)}$ matrices.

\subsubsection{Error Covariance Matrix for a Normal Response}
For a normally distributed response $\bsy_{(i)}$, the error covariance matrix $\bsR_{(i)}$ may take on a number of possible structures. \citet{karlem} provide the calculations for an unstructured covariance matrix in the presence of incomplete data.  We demonstrate the calculations of an AR(1) matrix assuming complete data, $\bsy$. For this single response, we will drop the subscript $(i)$ from the model terms defined previously. Suppose that $\bsy\mid\bse\sim N_n(\bsX\bsbeta+\bsZ\bse,\bsR)$ and the $(k,d)$ element of $\bsR$ is $\sigma^2\rho^{|k-d|}$. We will derive the score equations for $\sigma^2$ and $\rho$. The solution $S(\sigma^2,\rho)=\bds{0}$ may then be obtained by Newton-Raphson.
\small
\begin{align}
S\left(\rho\right)&=\int\der{\rho}\left[\log\left(\left|\bsR\right|^{-1/2}\right)-\frac{1}{2}\left(\bsy-\bsX\bsbeta-\bsZ\bse\right)^{\prime}\bsR^{-1}\left(\bsy-\bsX\bsbeta-\bsZ\bse\right)\right]f(\bse|\bsy;\moparm)\ud \bse\nonumber\\
&=-\frac{1}{2}\int\left[\te{tr}\left(\bsR^{-1}\dert{\rho}{\bsR}\right)-\left(\bsy-\bsX\bsbeta-\bsZ\bse\right)^{\prime}\bsR^{-1}\dert{\rho}{\bsR}\bsR^{-1}\left(\bsy-\bsX\bsbeta-\bsZ\bse\right)\right]f(\bse|\bsy;\moparm)\ud \bse\nonumber\\
&=-\frac{1}{2}\left[\te{tr}\left(\bsR^{-1}\dert{\rho}{\bsR}\right)-\left(\bsy-\bsX\bsbeta-\bsZ\widetilde{\bse}\right)^{\prime}\bsR^{-1}\dert{\rho}{\bsR}\bsR^{-1}\left(\bsy-\bsX\bsbeta-\bsZ\widetilde{\bse}\right)-\te{tr}\left(\bsZ^{\prime}\bsR^{-1}\dert{\rho}{\bsR}\bsR^{-1}\bsZ\widetilde{\bsv}\right)\right]\label{eq:scorerho}
\end{align}
\normalsize
For some patterned matrices or for data sets with incomplete data, $\bsR^{-1}$ must be calculated numerically. Furthermore, when some elements of $\bsy$ are missing, the score function is the sum of values calculated from (\ref{eq:scorerho}) over each dropout pattern in the data set \citep{karlem}. However, for a complete AR(1) matrix $\bsR$, $\bsR^{-1}$ has a simple tridiagonal expression \citep{demidenko} and thus \[\te{tr}\left(\bsR^{-1}\dert{\rho}{\bsR}\right)=\frac{-2\rho(n-1)}{1-\rho^2}.\]
Likewise,
\begin{align*}
S\left(\sigma^2\right)=&-\frac{1}{2}\left[\frac{n}{\sigma^2}-\left(\bsy-\bsX\bsbeta-\bsZ\widetilde{\bse}\right)^{\prime}\bsR^{-1}\dert{\sigma^2}{\bsR}\bsR^{-1}\left(\bsy-\bsX\bsbeta-\bsZ\widetilde{\bse}\right)-\te{tr}\left(\bsZ^{\prime}\bsR^{-1}\dert{\sigma^2}{\bsR}\bsR^{-1}\bsZ\widetilde{\bsv}\right)\right]
\end{align*}
Finally, if $\bsR=\sigma^2 I_n$, then
\begin{equation}\label{eq:sigM}
\widehat{\sigma}^2=\frac{1}{n}\left[\left(\bsy-\bsX\bsbeta-\bsZ\widetilde{\bse}\right)^{\prime}\left(\bsy-\bsX\bsbeta-\bsZ\widetilde{\bse}\right)+\te{tr}\left(\bsZ^{\prime}\bsZ\widetilde{\bsv}\right)\right] .
\end{equation}

\subsection{The E step}

Calculation of the components of the M step score vector requires the conditional mean, $\widetilde{\bse}$, and the conditional variance, $\widetilde{\bsv}$, of $f\left(\bse|\bsy_{(1)},\ldots,\bsy_{(q)};\moparm\right)$, as well as the conditional expectations appearing in (\ref{eq:scmumiss}). Letting $\left\{E\left[H(\bse)\right]\right\}_k$ denote the $(k)$-th component of the vector $E\left[H(\bse)\right]$, the M step updates require
\begin{align}
\left\{\te{E}\left[H(\bse)|\bsy_{(1)},\ldots,\bsy_{(q)};\moparm\right]\right\}_k&=\int \left\{H(\bse)\right\}_k f(\bse|\bsy_{(1)},\ldots,\bsy_{(q)};\moparm)\ud \bse\nonumber
\end{align}
for all $k$, where $H(\cdot)$ is a function of the random effects, and $\moparm$ is fixed at its value from the previous iteration. The M step updates for  $\bsbeta_{(i)},\bsG$ and $\bsR$ require $H(\bse)=\bse$ and $\widetilde{\bsv}=\te{var}[\bse|\bsy_{(1)},\ldots,\bsy_{(q)};\moparm].$
When response variable $i$ is non-normally distributed, the M step update for  $\bsbeta_{(i)}$  requires \[H(\bse)=\dert{\eta_{l(i)}}{\log[f(\bsy_{(i)}|\bse)]}\] for all $l\in \{1,\ldots,n_{(i)}\}$.
To solve these high-dimensional integration problems, we follow the examples of \citet{steele} and \citet{riz09} and use the fully exponential Laplace approximation of \citet{tierney89}, approximating the cumulant-generating function $\log\left[E\left(\exp\left[\bds{c}^{\prime}H(\bse)\right]\right)\right]$ at the mode $\widehat{\bse}=\widehat{\bse}^{(\bds{0})}$, where \[\widehat{\bse}^{(\bds{c})} = \te{argmax}_{\bse}\left\{\log\left[f\left(\bsy_{(1)},\ldots,\bsy_{(q)},\bse\right)\right] + \bds{c}^{\prime}H\left(\bse\right)\right\}.\]
The mode $\widehat{\bse}$ is obtained by Newton-Raphson, stopping when ${\bds{L}^{\prime}\bsSigma^{-1}\bds{L}<\alpha}$:
\begin{displaymath}
\widehat{\bse}^{\te{p}+1}=\widehat{\bse}^{\te{p}}-\left(\bsSigma^{\te{p}}\right)^{-1}\bds{L}\left(\widehat{\bse}^{\te{p}}\right)
\end{displaymath}
where superscript ``p'' is the iteration counter. Via matrix calculus \citep{matdif,har},
\begin{align}
\bds{L}(\bse)&=-\der{\bse}\left\{\left[\sum_{i=1}^q\log f\left(\bsy_{(i)}|\bse; \moparm\right)\right] +\log\left\{f\left(\bse\right)\right\}+\bsc^{\prime}H(\bse)\right\}|_{\bsc=\bds{0}}\nonumber\\
&=-\sum_{i=1}^q\sum_{l=1}^{n_{(i)}}\dert{\eta_{l(i)}}{\log[f(\bsy_{(i)}|\bse)]}\bsz_{l(i)}+\bsG^{-1}\bse \label{eq:gr.eta}
\end{align}
and $\bsSigma^{\te{p}}=\bsSigma^{(\bsc)}|_{(\bsc,\bse)=(\bds{0},\widehat{\bse}^{\te{p}})}$, with
\begin{align}
\bsSigma^{(\bsc)}&=-\frac{\partial^2}{\partial\bse\partial\bse^{\prime}}\left\{\left[\sum_{i=1}^q\log f\left(\bsy_{(i)}|\bse; \moparm\right)\right] +\log\left\{f\left(\bse\right)\right\}+\bsc^{\prime}H(\bse)\right\}\nonumber\\
&=-\sum_{i=1}^q\sum_{l=1}^{n_{(i)}}\sum_{m=1}^{n_{(i)}}\frac{\partial^2\log[f(\bsy_{(i)}|\bse)]}{\partial\eta_{l(i)}\partial\eta_{m(i)}}\bsz_{l(i)}\bsz_{m(i)}^{\prime}+\bsG^{-1}-\frac{\partial^2\bsc^{\prime}H(\bse)}{\partial\bse\partial\bse^{\prime}}. \label{eq:sigma}
\end{align}
Equivalently, with $\boldsymbol\eta_{(i)}=\bsX_{(i)}\bsbeta_{(i)}+\bsZ_{(i)}\bse$ and
$\bsW_{(i)}=-\nabla^2_{\boldsymbol\eta_{(i)}}\log f(\bsy_{(i)}|\bse)$,
\[
\bsSigma^{(\bsc)}=\bsG^{-1}+\sum_{i=1}^q\bsZ_{(i)}^{\prime}\bsW_{(i)}\bsZ_{(i)}
-\frac{\partial^2\{\bsc^{\prime}H(\bse)\}}{\partial\bse\partial\bse^{\prime}}.
\]
The derivatives are evaluated at the current iterate during mode finding and at $\widehat{\bse}^{(\bsc)}$ in the Laplace approximation. For conditionally independent observations $\bsW_{(i)}$ is diagonal, giving the original single-sum expression. For a normal response with residual covariance $\bsR_{(i)}$, $\bsW_{(i)}=\bsR_{(i)}^{-1}$; the full matrix is required even when it is not diagonal. Appendix A gives this calculation explicitly. Correlation among random effects in $\bsG$ does not itself require mixed predictor derivatives.

For large problems, the algorithm may be sped up by performing only a single NR step in early iterations and then running the algorithm to convergence in later iterations. Although the Laplace approximation used in the E step requires finding the conditional mode, this process requires more NR iterations in the first few E steps as the parameter estimates begin to stabilize. The resulting parameter estimates, $\moparm^{(k)}$, after $k$ such abbreviated EM iterations may be viewed as initial conditions for the full EM algorithm. As with other iterative likelihood methods, a stationary solution need not be the global maximum, and sensitivity to initial values should be explored in each application.

  Once the NR algorithm converges to the mode $\widehat{\bse}$, the next step is to apply a fully exponential Laplace approximation to $E[\exp\{\bsc^{\prime}H(\bse)\}]$. We apply the result of Theorem 2 of \citet{tierney89}. Using properties of the cumulant-generating function,
\begin{align}
\left\{E\left[H(\bse)|\bsy_{(1)},\ldots,\bsy_{(q)};\moparm\right]\right\}_k&\approx\der{c_k}\left.\left\{\bsc^{\prime}H\left(\widehat{\bse}^{(\bsc)}\right)+\log\left[\det\left(\bsSigma^{(\bsc)}\right)^{-1/2}\right]\right\}\right|_{\bsc=\bds{0}}\nonumber\\
&=\bep{k}H(\widehat{\bse})-\frac{1}{2}\tr\left(\left.\bsSigma^{-1}\left\{\frac{\partial\bsSigma^{(c)}}{\partial c_k}\right|_{(\bds{c},\bse)=(\bds{0},\widehat{\bse})}\right\}\right), \label{eq:etahat}
\end{align}
where $\be{k}$ is the vector of zeros with a 1 in the $k$-th component. The $(k,d)$ element of $\var(\bse)$, evaluated at $(\bds{c},\bse)=(\bds{0},\widehat{\bse})$, is
\begin{align}
\left\{\var\left(\bse|\bsy_{(1)},\ldots,\bsy_{(q)};\moparm\right)\right\}_{kd}&\approx\bep{k}\bsSigma^{-1}\be{d}-\frac{1}{2}\tr\left(\bsSigma^{-1}\frac{\partial^2\bsSigma}{\partial c_k \partial c_d}-\bsSigma^{-1}\frac{\partial\bsSigma}{\partial c_d}\bsSigma^{-1}\frac{\partial\bsSigma}{\partial c_k}\right) \label{eq:varetahat}.
\end{align}

The first-order Laplace approximation consists of only the first terms of (\ref{eq:etahat}) and (\ref{eq:varetahat}) \citep{kass89}. The terms involving the trace in both equations are the fully exponential corrections to the first-order Laplace approximation. As discussed in Section \ref{sec:computation}, for problems where the fully exponential corrections are too expensive to calculate, the first-order approximation may be used instead. Notice that the first-order Laplace approximation only uses the first two derivatives of the conditional log-likelihood. We will see that the fully exponential corrections in (\ref{eq:etahat}) and (\ref{eq:varetahat}) require the third and fourth derivatives, respectively. This suggests that the EBLUPs $(\widetilde{\bse})$ from GLMMs using different link functions for the same distribution will show greater differences when the fully exponential corrections are applied.

Calculations of the fully exponential corrections are furnished in Appendix B. These are generalizations of the calculations presented by \citet{riz09}. The non-nested structure of our model affects the terms derived inside of the trace functions.

\subsection{Convergence and EM Standard Errors}
For the exact EM algorithm, suitable regularity conditions imply that limit points are stationary points of the observed data likelihood \citep{wu83}. Compactness conditions can sometimes be enforced by restricting the parameter space \citep{mcculloch94,demidenko}. These results do not automatically extend to the Laplace-approximated E steps used here; continuity of the updates alone does not establish likelihood ascent or convergence. Convergence should therefore be assessed numerically, with sensitivity to starting values and approximation order examined. The approximation error is discussed in Section~\ref{sec:error}.

EM algorithms have been documented to converge slowly when there is a large number of random effects, or when a variance component is near zero. \citet{lindstrom88} note how the EM algorithm is often outperformed by a Newton-Raphson routine when fitting mixed models. However, NR algorithms tend to step outside of the parameter space when the $\bsG$ matrix is nearly singular. \citet{karlem} show how the NR routine of SAS GLIMMIX fails in the presence of nearly perfectly correlated random effects, while an EM algorithm converges successfully since the estimated $\bsG$ matrix is guaranteed to be positive definite after each iteration. It is feasible that applications of the joint model presented in Section~\ref{sec:cre} will have highly correlated random effects across response variables, especially in situations when assumptions made by shared parameter models \citep{wu88} are valid. \citet{riz09} demonstrate how an EM approach may be converted to a quasi-Newton approach, which could be useful in the absence of strong correlations.

The observed data score vector $S(\moparm)$ is equal to the conditional expectation of the complete data score vector when the conditional distribution is evaluated at the same $\moparm$ \citep{louis}. As proposed by \citet{jam00}, it is possible to calculate the observed information matrix at the MLE $\widehat{\moparm}$ using a central difference approximation to the Hessian
\begin{align*}
-\left.\partial S(\moparm)/\partial \moparm \right|_{\moparm=\widehat{\moparm}}.
\end{align*}

\subsection{Approximation Error} \label{sec:error}
In this section we discuss the degree of error in the fully exponential Laplace approximation presented in Section~\ref{sec:estimation}. Theorem 1 of \citet{tierney89} demonstrates that, for the approximations appearing in (\ref{eq:etahat}) and (\ref{eq:varetahat}),
\begin{align*}
E\left[H\right]&=\hat{E}\left[H\right]+ O(r^{-2})\\
V\left[H\right]&=\hat{V}\left[H\right]+ O(r^{-3})
\end{align*}
where the hat denotes the fully exponential Laplace approximation, and $r=r(\bsy_{(1)},\ldots,\bsy_{(q)})$ is a measure of the size of the data set $(\bsy_{(1)},\ldots,\bsy_{(q)})$ such that $r\rightarrow\infty$ as the size of the data set grows \citep{evans95}.
Applying the Laplace approximation to a non-normal model $f(\bsy_{(i)}|\bse)$ will result in approximation error because the $\bse_{(i)}$ enter the integrand via a nonlinear link function. The approximation error depends on the amount of information in $\bsy_{(i)}$ and the random effects structure of $f(\bsy_{(i)}|\bse)$. If the random effects structure is nested, then $r$ equals the minimum number of observations on any of the components of $\bse_{(i)}$, along the lines of the work by \citet{vonesh96}. A numerical study of the performance of the first-order Laplace approximation in nested binary and Poisson models is provided by \citet{joe08}.

Models for normally distributed responses do not contribute to the approximation error. For illustration, consider the joint modeling of a normal response $\bsy_{(1)}$ with a non-normal response $\bsy_{(2)}$. Applying the Laplace approximation to the marginal likelihood,
\begin{align}
&{}\iint f(\bsy_{(1)}|\bse_{(1)})f(\bsy_{(2)}|\bse_{(2)})f(\bse_{(1)},\bse_{(2)}) \ud\bse_{(1)}\ud\bse_{(2)}\nonumber\\
=&{}\int f(\bsy_{(2)}|\bse_{(2)})\left\{\int f(\bsy_{(1)}|\bse_{(1)})f(\bse_{(1)},\bse_{(2)})\ud\bse_{(1)}\right\}\ud\bse_{(2)}\nonumber\\
=&{}\int f(\bsy_{(2)}|\bse_{(2)}) I(\bse_{(2)};\bsy_{(1)}) \ud\bse_{(2)} . \label{eq:mislaplace}
\end{align}
The Laplace approximation is exact for $\int f(\bsy_{(1)}|\bse_{(1)})f(\bse_{(1)},\bse_{(2)})\ud\bse_{(1)}$, since $\bse_{(1)}$ enters the model linearly \citep{pinheiro95}. The result, $I(\bse_{(2)};\bsy_{(1)})=f(\bsy_{(1)})f(\bse_{(2)}\mid\bsy_{(1)})$, is proportional to the density of a normal distribution with mean and variance depending on the covariance between $\bse_{(1)}$ and $\bse_{(2)}$. For example, if these effects are uncorrelated -- and hence independent, due to their joint normality -- then $I(\bse_{(2)};\bsy_{(1)})=f(\bse_{(2)})f(\bsy_{(1)})$.

If the number of random effects in the non-normal model $f(\bsy_{(i)}|\bse_{(i)})$ increases with the sample size, then the dimension of the approximated integrals in the E step also increases with the sample size. This property is not typical for applications of the Laplace approximation; \citet{shun95} describe how a modification to the first-order Laplace approximation is needed to retain its usual order of convergence, although they did not study the behavior of the fully exponential approximation, and note that it may not suffer the same extent of deterioration in this setting as the first-order approximation. The results of \citet{shun97} show, in an application to the salamander mating data \citep{mccullagh89}, that the uncorrected first-order Laplace approximation outperforms PQL. This is reasonable since PQL makes an additional approximation to one of the terms in the first-order Laplace approximation. These results motivate the use of first-order and fully exponential Laplace approximations as alternatives to PQL (and to PL when the scale parameter is fixed), but do not guarantee an improvement in every setting or establish consistency when the dimension increases with the sample size.

\section{Computation}\label{sec:computation}
In order to compute the fully exponential approximation in a reasonable amount of time, we use a few key facts. First, the operations inside of the trace function in (\ref{eq:varetahat}) contain some elements that are common to all of the components of $\widetilde{\bsv}$ and do not need to be re-calculated for each component. Secondly, some of the needed computations involve multiplication by sparse matrices, greatly reducing the number of needed arithmetic operations. Finally, not all of the components of $\widetilde{\bsv}$ will be used in the M step: the unused components may be ignored.

To see why only some of the components of $\widetilde{\bsv}$ are needed, note that the M step update for $\bsbeta_{(i)}$ does not use $\widetilde{\bsv}$ and observe how the covariance matrix $\widetilde{\bsv}$ is used in the M step update of $\bsGamma$ in (\ref{eq:gamM}). The M step update of $\bsGamma$ requires only relatively small block-diagonal portions of  $\widetilde{\bsv}$. In the context of the Poisson-binary sports ranking model, these are $3\times 3$ blocks, corresponding to the non-zero components of $\bsG$. As a result, only $126\times 6=756$ of the $378\times379/2=71{,}631$ components in the upper triangle of $\widetilde{\bsv}$ will be used in the application of Section~\ref{sec:sports}.

\begin{figure}
\caption{The shaded pixels (including the blocks on the diagonal) indicate which components of $\widetilde{\bsv}$ need fully exponential corrections from  (\ref{eq:varetahat}) in the application of Section~\ref{sec:apmodel}.}
\label{plot:comp_list}
\centering
\includegraphics[scale=.5]{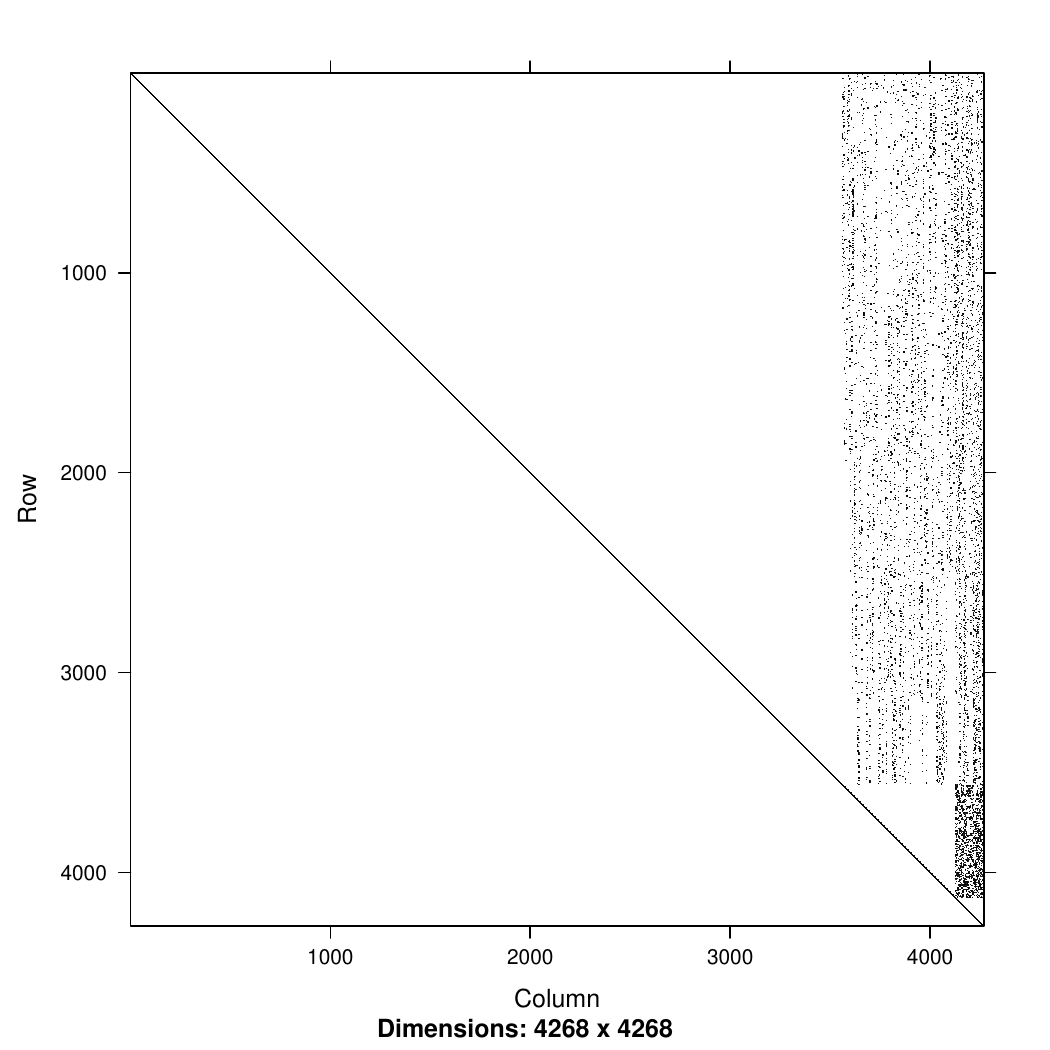}
\end{figure}

For a normally distributed response $i$ with a diagonal error covariance matrix, the M step update of  $\sigma^2$ requires
\begin{equation}\label{eq:had}
\te{tr}\left[\bsZ_{(i)}^{\prime}\bsZ_{(i)}\widetilde{\bsv}\right]= \sum_{\te{components}}\left[\left(\bsZ_{(i)}^{\prime}\bsZ_{(i)}\right)\circ\widetilde{\bsv}\right]
\end{equation}
where $\circ$ represents the Hadamard product. Thus, the only components of $\widetilde{\bsv}$ that are needed in addition to the aforementioned block-diagonal elements are those that have the same indices as non-zero components of the sparse matrix $\bsZ_{(i)}^{\prime}\bsZ_{(i)}$. The plot in Figure \ref{plot:comp_list} indicates which components of $\widetilde{\bsv}$ require corrections in the application of Section~\ref{sec:apmodel}. Only $14138/9110046=0.0016$ of the elements in the upper triangle of $\widetilde{\bsv}$ need to be calculated. Unfortunately, this feature does not extend to normally distributed responses with a non-diagonal error covariance matrix, $\bsR$. As is apparent from the score function for $\rho$ in (\ref{eq:scorerho}), the updates for the components of $\bsR$ depend on
\[\sum_{\te{components}}\left[\left(\bsZ^{\prime}\bsR^{-1}\dert{\rho}{\bsR}\bsR^{-1}\bsZ\right)\circ\widetilde{\bsv}\right]\]
Although such models are capable of being estimated by the proposed method with a fully exponential Laplace approximation, the computations are not scalable. For these or other situations where the fully exponential corrections are too expensive to calculate, the first-order Laplace approximation produced by the EM algorithm will still be scalable. Or, as suggested in Section~\ref{sec:sim}, a partial fully exponential correction using only (\ref{eq:etahat}) may still be calculated efficiently. The main computational limitation on the first-order Laplace approximation is the memory required to hold $\bsSigma^{-1}$, which is dense. With $m$ representing the length of $\bsb$, this matrix requires approximately $8m^2$ bytes to store in R (half of that if it is explicitly stored as a symmetric matrix). One gigabyte is necessary for $m$=11,180, two gigabytes for $m$=15,811, three gigabytes for $m$=19,364, etc.

The computational burden of calculating the fully exponential corrections in (\ref{eq:varetahat}) is tremendous. The dimension of each of the matrices $\bsSigma^{-1}$, ${\partial\bsSigma}/{\partial c_k}$, and ${\partial^2\bsSigma}/(\partial c_k \partial c_d)$ is $m \times m$. Even ignoring the calculations required to obtain the derivatives ${\partial\bsSigma}/{\partial c_k}$ and ${\partial^2\bsSigma}/(\partial c_k \partial c_d)$, the term inside the trace function of (\ref{eq:varetahat}) requires approximately $4(2m^3-m^2)$ calculations. This results in a requirement of just over $6.2\times 10^{11}$ floating point operations to calculate the fully exponential correction for a single component of $\widetilde{\bsv}$ for the example in Section~\ref{sec:apmodel}, if implemented naively. Since, in this example, there are $9.1$ million components in the upper-triangle of the symmetric $\widetilde{\bsv}$ matrix, this implementation would require around $5.7\times 10^{18}$ operations per iteration, excluding the operations needed to calculate the derivatives. The computer used for this application runs at 7.2 billion floating point operations per second, and would require around 25 years to execute one iteration of a naively implemented fully exponential correction to the full $\widetilde{\bsv}$ matrix. We reduce this time to 27 hours by calculating only the necessary corrections and by using properties of the trace such as the one shown in (\ref{eq:had}).

For smaller problems, it would be possible to calculate the derivatives  ${\partial\bsSigma}/{\partial c_d}$, $d=1,\ldots,m$ and save them in memory, greatly speeding up the fully exponential approximation. However, this would require too much memory for most applications. As a compromise, we first calculate and store ${\partial\bsSigma}/{\partial c_1}$. We then calculate all of the corrections that require this matrix, and then repeat for $c_2,\ldots,c_m$. The loop calculating the corrections for these components is an excellent candidate for parallelization, which could be done using one of the packages for parallel computing described in \citet{eugster}.

The simulations in Section~\ref{sec:sim} suggest that using the corrections to $\widetilde{\bse}$ inside the trace function of (\ref{eq:etahat}), but not those to $\widetilde{\bsv}$ in the trace function of (\ref{eq:varetahat}), produces results that are relatively close to those obtained by the fully exponential approximations while requiring only a fraction of the time. This seems reasonable since the M step updates for the $\bsbeta_{(i)}$ are functions of $\widetilde{\bse}$, but not $\widetilde{\bsv}$. Furthermore, the M step update for $\bsG$ is composed of diagonal blocks from the matrix $\widetilde{\bsv}+\widetilde{\bse}\widetilde{\bse}^{\prime}$. Hence $\widehat{\bsG}$ does experience at least part of the benefit of the fully exponential corrections when only the corrections to $\widetilde{\bse}$ are included. We take advantage of this by first running the first-order Laplace approximation to convergence, then introducing the corrections in the trace function of (\ref{eq:etahat}) and again running to convergence. We then treat the resulting parameter estimates as initial values for the fully exponential approximation (using the corrections from the trace functions of both (\ref{eq:etahat}) and (\ref{eq:varetahat})). This minimizes the number of iterations that require corrections to $\widetilde{\bsv}$, which is important given the tremendous amount of time taken to calculate these corrections, as seen in Table \ref{tab:times}.

We rely heavily on the sparse matrix methods of the Matrix package \citep{Matrix}. To illustrate, only 13534 components of the $5697 \times 4268$ $\bsZ_{(1)}$ matrix and 3557 components of the $3557\times 4268$ $\bsZ_{(2)}$ matrix for the application in Section~\ref{sec:apmodel} are nonzero. This has important implications for the program: for example, the calculation of $\bsZ_{(1)}^{\prime}\bsZ_{(1)}$ in R registers a time of 0 seconds when $\bsZ_{(1)}$ is stored as a sparse matrix (requiring 0.18 MB), and 57 seconds when it is treated as dense (requiring 194.52 MB). \citet{karlem} describe how SAS runs out of memory when attempting to fit a multiple membership linear mixed model with around 3500 random effects on a machine with 4 GB of RAM.

\section{Applications}\label{sec:application}
Two applications are presented in this section to illustrate the computational methods of this paper.

\subsection{Sports Ranking Model}\label{sec:sports}
The first application fits the Poisson-binary model in Section~\ref{ssec:ME} to the regular season data from the 2012 NCAA FBS Division of college football. This data set contains results from 844 games including 126 teams (lower-division teams are consolidated into a single effect, labeled ``fcs''). Table~\ref{tab:sportstimes} summarizes the median time required per iteration for the first-order Laplace approximation, fully exponential corrections from (\ref{eq:etahat}) only, and the complete fully exponential Laplace approximation.

\begin{figure}
\caption{College football defensive vs. offensive ratings at the end of the regular season. The size of the plotted point increases with the win propensity rating of each team.}
\label{plot:f2012}
\centering
\includegraphics[width=15cm]{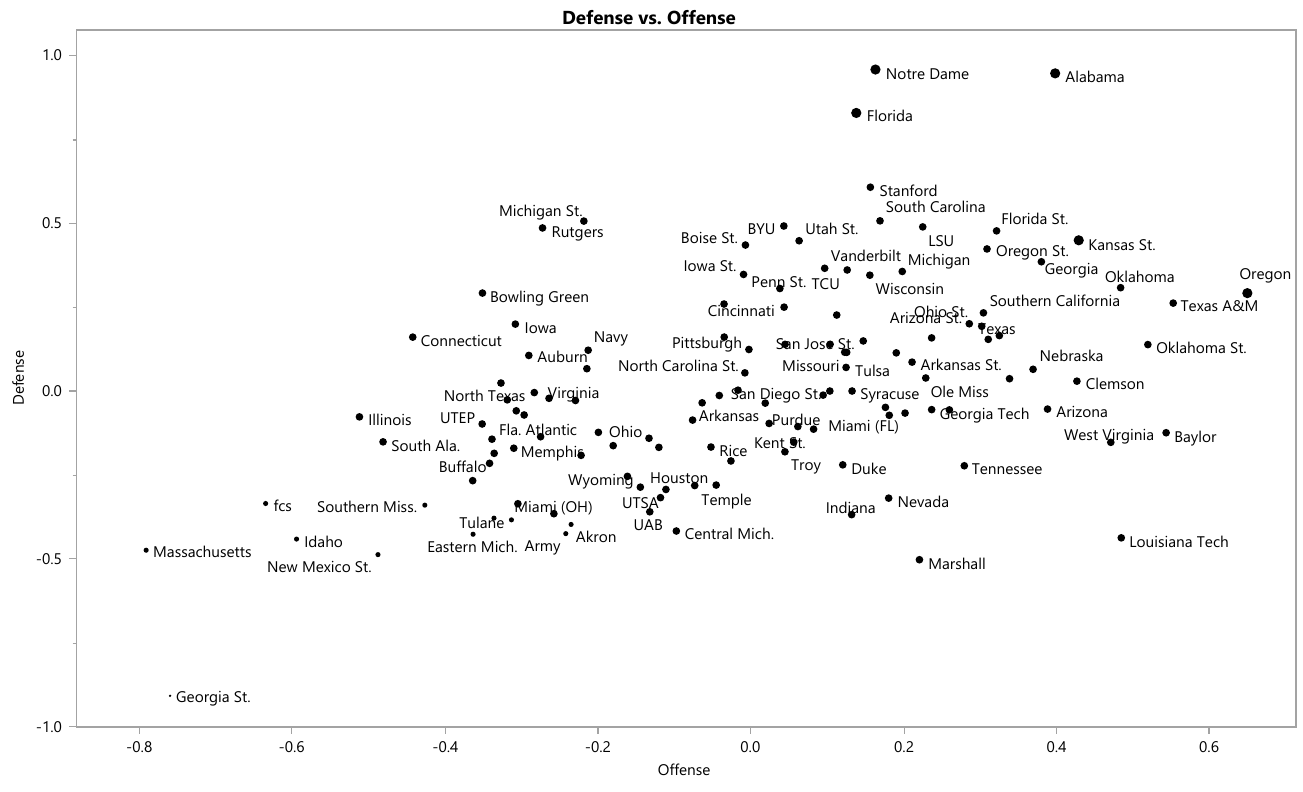}
\end{figure}

The EBLUPs for the offensive and defensive ratings are plotted in Figure~\ref{plot:f2012}. The size of each plotted point is proportional to the EBLUP for the team's win propensity. Table~\ref{tab:gmats} presents the estimated $\bsG$ covariance matrices from the first-order Laplace, first-order Laplace with corrections from (\ref{eq:etahat}), and fully exponential Laplace approximations. The offensive and defensive ratings are positively correlated, suggesting teams with good (bad) offenses tend to also have good (bad) defenses. Furthermore, it appears that these ratings are equally correlated with the win propensity. Notice how the estimates for the variance components increase with the quality of the approximation, and how the estimated components using the correction from (\ref{eq:etahat}) are relatively close to those from the fully exponential approximation.

 Fitting the Poisson and binary models separately leads to the estimates in the second row of Table~\ref{tab:gmats}. The most noticeable difference is that the variance of the win propensity rating increases in the joint model. To a lesser extent, the variances of the offensive and defensive ratings increase as well.

\begin{table}
\caption{Estimated $\bsG$ covariance matrices from the joint Poisson-binary model for scores and win/loss indicators (first row) and from independent models for the scores and indicators (second row).}
\label{tab:gmats}
		\begin{tabular}{ccc}
		\toprule
		First-order&FE: (\ref{eq:etahat}) only&Fully Exponential (FE)\\
		\midrule
$\begin{pmatrix}
0.0917	&0.0472	&0.3151\\
0.0472	&0.1022	&0.3355\\
0.3151	&0.3355	&1.4732
\end{pmatrix}$
&$\begin{pmatrix}
0.0920  &0.0473    &     0.3215\\
0.0473  &0.1025    &     0.3425\\
0.3215  &0.3425    &     1.5297
\end{pmatrix}$
&$\begin{pmatrix}
  0.0920 & 0.0473     &    0.3216\\
  0.0473 & 0.1025     &    0.3425\\
  0.3216 & 0.3425     &    1.5306
\end{pmatrix}$\\
\addlinespace[1.5ex]
$\begin{pmatrix}
0.0908&  0.0460  &     0\\
  0.0460&   0.1015&     0\\
0 &0    & 0.7614
\end{pmatrix}$
&$\begin{pmatrix}
0.0908& 0.0460    &     0\\
 0.0460  &   0.1015&     0\\
0 &0    &  1.1105
\end{pmatrix}$
&$\begin{pmatrix}
0.0908& 0.0460    &     0\\
 0.0460  &   0.1015&     0\\
0 &0    & 1.1407
\end{pmatrix}$\\
\bottomrule
		\end{tabular}
\end{table}

\begin{table}
  \centering
  \caption{Median time per iteration (seconds) for the 2012 college football data.}
    \begin{tabular}{rlll}
    \toprule
      Approximation    & Binary & Poisson  & Poisson-binary \\
    \midrule
    First-order & 0.11  & 0.80 & 1.20 \\
    FE: (\ref{eq:etahat}) only     & 1.07 & 4.32  & 8.72 \\
    Fully Exponential (FE) & 1.79 & 11.33 & 33.90 \\
    \bottomrule
    \end{tabular}%
  \label{tab:sportstimes}%
\end{table}%

\subsection{Model for Continuous Outcome with Missing Data Indicator}\label{sec:apmodel}
Longitudinal studies commonly finish with missing observations: ignoring them requires an assumption that the observations are missing at random (MAR). When the missingness process at time $t$ depends on the values that would be observed at time $t$, or on latent variables contributing to the value of the observation at time $t$, the missing observations are missing not at random (MNAR). In this situation, obtaining the appropriate parameter estimates and standard errors requires consideration of a model for the missingness mechanism \citep{little87}.

\citet{karlcpm} develop a correlated random effects model to explore the sensitivity of the EBLUPs from the ``generalized persistence'' (GP) value-added model (VAM) \citep{mariano10} to assumptions about the missing data. They present an example using calculus 2 and 3 grades from a large public university, with 5697 grades recorded for 3557 students from 189 and 144 calculus 2 and 3 courses, respectively. In the notation of Section~\ref{sssec:gmat}, $U=3$, $q=2$, $V_1=3$, $M_1=189$, $V_2=1$, $M_2=144$, $V_3=1$, $M_3=3557$.
In the complete-data GP VAM \citep{mariano10}, each calculus 2 teacher has an effect on their students' calculus 2 grades and an effect on their students' calculus 3 grades. Calculus 3 teachers have an effect associated with their students' calculus 3 grades. The two effects of the calculus 2 teachers are allowed to be correlated via the $2\times 2$ covariance matrix $\bsGamma_{teach,1}$. These are assumed to be independent from the effects of the calculus 3 teachers, whose effects are modeled by the variance component $\Gamma_{teach,2}$. Furthermore, a random intercept for each student, with associated variance component $\Gamma_{stu}$, allows the student grades for calculus 2 and calculus 3 to be correlated.
All of the students completed calculus 2, but not all of the students completed calculus 3. To study the sensitivity of the random teacher effects from the GP VAM to the assumption that missing calculus 3 scores are MAR, \citet{karlcpm} build a model for student completion of calculus 3 (the ``attendance model'').  In their joint model referred to as MNAR-t, student attendance in calculus 3 is modeled as a function of effects associated with their calculus 2 teachers, recognizing that a teacher's effect on grades may be associated with the propensity of his or her future students to complete calculus 3. The attendance effect of each calculus 2 teacher is allowed to be correlated with the teacher's two effects from the test score model, and these effects are thus modeled with a $3 \times 3$ unstructured covariance matrix $\bsGamma_{teach,1}^*$ in the joint model. The student grades, $\bsy_{(1)}$, are assumed to be normally distributed and conditionally independent: the mixing of classmates from year to year leads to a multiple membership structure in $\bsZ_{(1)}$. The binary calculus 3 attendance indicators, $\bsy_{(2)}$, are modeled with a probit link. The $\bsG$ matrix for this example takes the form
\begin{equation*}
\bsG=\te{cov}(\bse)=\te{blockdiag}\left(\Gamma_{\te{stu}},\ldots,\Gamma_{\te{stu}},\bsGamma_{teach,1}^*,\ldots,\bsGamma_{teach,1}^*,\ldots,\Gamma_{teach,2},\ldots,\Gamma_{teach,2}\right)
\end{equation*}
We refer to \citet{karlcpm} for a listing of parameter estimates and a discussion of the results. Table~\ref{tab:times} lists the median iteration times for different approximations for this model.

\begin{table}[htbp]
  \centering
  \caption{Median run times for iterations with first-order Laplace approximation, fully exponential corrections from (\ref{eq:etahat}) only, and fully exponential Laplace approximation.}
    \begin{tabular}{rl}
    \addlinespace
    \toprule
    Approximation & Iteration Time  \\
    \midrule
    First-order& 1.5 min \\
    FE:(\ref{eq:etahat})-only & 3.0 min  \\
    Fully Exponential (FE) & 27 hr \\
    \bottomrule
    \end{tabular}%
  \label{tab:times}%
\end{table}%

\section{Simulations}\label{sec:sim}
We illustrate the performance of our algorithm with two sets of simulations based on the binary model for the home team win/loss indicators. We focus on the binary model to highlight the performance gains available from the Laplace and fully exponential Laplace approximations in a non-nested application. \citet{joe08} runs similar trials for nested binary and Poisson models. We have made both our simulation code and our code for the Poisson-binary model available in the online supplement, providing readers with the necessary tools to run more extensive simulations.
\begin{align}\label{eq:binary2}
f(\bsy_{(2)}|\bse)=&\prod_{g=1}^n \left[\Phi\left\{\left(-1\right)^{1-y_{g(2)}}\left(\beta_{(2)}+\bsz_{g(2)}^{\prime}\bsb\right)\right\}\right]\\
\bse\sim&N(\bds{0},\sigma^2_w\te{\bf{I}})
\end{align}
The only difference between this model and the one specified in (\ref{eq:binary}) is the presence of the constant fixed effect, $\beta_{(2)}$. Since the $\bsy_{(2)}$ are indicators for a home-win, $\beta_{(2)}$ models the home-field advantage on the probit scale.

 Each simulation consists of 500 runs with a vector of 100 ``team ratings.'' In one simulation, the ratings are generated from a $N(0,0.5)$ distribution, matching the structure assumed by the model. In the other simulation, the ratings are simulated from a $t_3$ distribution, and scaled by $1/\sqrt{6}$ to have variance 0.5. Each ``team'' plays a random schedule of 4 home games and 4 away games. The binary win/loss outcomes are evaluated with model (\ref{eq:binary2}). A home field effect of 0.1 was built into the simulations, meaning, for example,  the home team in a game between two evenly matched teams has a $\Phi(0.1)\approx54.0\%$ chance of winning. We compare the performance of a pseudo-likelihood routine (using residual maximum likelihood), the first-order Laplace approximation, the first-order Laplace approximation with fully exponential corrections from Equation~(\ref{eq:etahat}) only, and the fully exponential Laplace approximation. Our PL routine gives the same results that would be obtained by the default METHOD=RSPL setting of SAS PROC GLIMMIX. We have provided a copy of this program in the supplementary online material to accompany the Laplace routines.
\begin{table}[htbp]
  \centering
  \caption{Median parameter estimates from 500 runs, using a $N(0,0.5)$ distribution.}
    \begin{tabular}{rrrrrr}
    \toprule
     & Simulated Value & RSPL  & Laplace & FE:(\ref{eq:etahat})-only & FE Laplace \\
    \midrule
    $\sigma^2_w$& 0.5   & 0.309 & 0.335&0.479 & 0.506 \\
    $\beta_{(2)}$ & 0.1   & 0.084 & 0.092 & 0.098& 0.100 \\
    \bottomrule
    \end{tabular}%
  \label{tab:simtabn}%
\end{table}%
\begin{table}[htbp]
  \centering
  \caption{Median parameter estimates from 500 runs, using a $\frac{1}{\sqrt{6}}t_3$  distribution.}
    \begin{tabular}{rrrrrr}
    \toprule
     & Simulated Value & RSPL  & Laplace & FE:(\ref{eq:etahat})-only&FE Laplace \\
    \midrule
   $\sigma^2_w$ & 0.5   & 0.207 & 0.224 &0.295 &0.310 \\
   $\beta_{(2)}$ & 0.1   & 0.088 & 0.094 &0.097 &0.098 \\
    \bottomrule
    \end{tabular}%
  \label{tab:simtab}%
\end{table}%
\begin{table}[htbp]
  \centering
  \caption{Median run times (seconds) for iterations with first-order Laplace approximation, fully exponential corrections from (\ref{eq:etahat}) only, and fully exponential Laplace approximation.}
    \begin{tabular}{rl}
    \addlinespace
    \toprule
    Approximation & Iteration Time  \\
    \midrule
    First-order& 0.25 \\
    FE:(\ref{eq:etahat})-only & 0.42 \\
    Fully Exponential (FE) & 1.45 \\
    \bottomrule
    \end{tabular}%
  \label{tab:times2}%
\end{table}%

The behavior of the parameter estimates in Tables \ref{tab:simtabn} and \ref{tab:simtab} is consistent with the findings of previous work \citep{tierney89,breslow95,lin96,steele}. The RSPL parameter estimates are biased, especially the variance component estimate. The first-order and fully exponential Laplace approximations provide progressive improvements. The estimates using FE:(\ref{eq:etahat})-only show much of the improvement of the fully exponential Laplace approximation in a fraction of the time.

\section{Summary}\label{sec:summary}

We demonstrate how the use of an EM algorithm with a fully exponential Laplace
approximation proposed by \citet{steele} and \citet{riz09} may be adapted for the
estimation of multiple response, non-nested, generalized linear mixed models. We illustrate the
utility of the method in the context of a correlated random effects model proposed by \citet{karlcpm} for joint modeling of student test scores and attendance, as well as with a Poisson-binary model for evaluating sports teams. We have provided sample code to build pseudo-likelihood, first-order Laplace, and fully exponential Laplace routines in the online supplementary material. Readers interested in modifying the R code for their own applications must primarily 1) construct the appropriate $\bsZ_{(i)}$ matrices for their model, 2) change the definitions of four functions which return the derivatives appearing in Section \ref{ssec:derivatives}, and 3) modify the M step updates for the $\bsG$ matrix.

The presence of a linear predictor in GLMMs leads to a simplification that allows for the development of a template for joint estimation of an arbitrary number of GLMMs with different types of responses. In fact, there is no requirement here that the response distribution belong to the exponential family. The template requires the relevant conditional log-likelihood derivatives and well-defined mode and curvature calculations. For dependent non-normal observations, mixed predictor derivatives are required in addition to the scalar derivatives listed here. The method could be extended to more general nonlinear mixed models; however, some simplifications, such as (\ref{eq:chain}), would no longer apply in this setting.

The improved numerical accuracy from the algorithms presented in this paper is important for
calculating the maximum likelihood estimates themselves, but comes into play even more when
the GLMM model-fitting is iterated in a jackknife or bootstrap.  \citet{lohr07} illustrates the
sensitivity of bootstrap and jackknife calculations of small area estimates to the numerical method used
to calculate parameter estimates of a GLMM. The methods in this paper, with their increased accuracy,
are potentially useful in conjunction with resampling methods for variance estimation.

The proposed EM algorithm and computational routine may also be effectively applied to nested models. \citet{callcenter} present a mixed model for forecasting the volume of calls received by a call center over the course of a day. The number of calls in each period is transformed to be approximately normal. Independently, the mean service time in each period is modeled using a normal distribution, and the predictions for call volume and service times are combined for an estimate of the workload faced by the center. It is feasible that the number of calls received in a period may be correlated with the length of those calls. The correlated random effects model described in this paper may be used to develop either a normal-exponential or a Poisson-exponential model for call center volume and service time.

\appendix
\section*{Appendix A: Derivatives of the Conditional Log-Likelihoods}
\subsection*{Poisson Response}
For a Poisson response with a log link,
\begin{align*}
\log f(\bsy_{(i)}|\bse)&=\sum_{l=1}^{n_{(i)}} \left\{-\log\left(y_{l(i)}!\right)+y_{l(i)}\eta_{l(i)}-\exp\left(\eta_{l(i)}\right)\right\}\\
\dert{\eta_{l(i)}}{\log[f(\bsy_{(i)}|\bse)]}&=y_{l(i)}-\exp\left(\eta_{l(i)}\right)\\
\dertwot{\eta_{l(i)}}{\log[f(\bsy_{(i)}|\bse)]}&=\derthreet{\eta_{l(i)}}{\log[f(\bsy_{(i)}|\bse)]}=\derfourt{\eta_{l(i)}}{\log[f(\bsy_{(i)}|\bse)]} = -\exp\left(\eta_{l(i)}\right)
\end{align*}
\subsection*{Binary Response}
Suppose a binary response $\bsy_{(i)}$ is modeled with a probit link, and let $\Phi(\cdot)$ represent the standard normal distribution function.
\begin{align*}
\log f(\bsy_{(i)}|\bse)&=\sum_{l=1}^{n_{(i)}} \left\{y_{l\left(i\right)}\log\left(\Phi\left[\eta_{l(i)}\right]\right)+\left(1-y_{l\left(i\right)}\right)\log\left(1-\Phi\left[\eta_{l\left(i\right)}\right]\right)\right\}\\
&=\sum_{l=1}^{n_{(i)}}\log\left(\Phi\left[\left(-1\right)^{1-y_{l(i)}}\eta_{l(i)}\right]\right)
\end{align*}
Let $\eta^*_{l(i)}=\left(-1\right)^{1-y_{l(i)}}\eta_{l(i)}$ to compress notation,
\begin{align*}
\dert{\eta_{l(i)}}{\log[f(\bsy_{(i)}|\bse)]}&=\left(-1\right)^{1-y_{l(i)}}\frac{\phi(\eta^*_{l(i)})}{\Phi(\eta^*_{l(i)})}\\
\dertwot{\eta_{l(i)}}{\log[f(\bsy_{(i)}|\bse)]}&=\frac{\frac{\partial\phi\left(\eta^*_{l(i)}\right)}{\partial\eta_{l(i)}}\Phi\left(\eta^*_{l(i)}\right)-\phi^2\left(\eta^*_{l(i)}\right)}{\Phi^2\left(\eta^*_{l(i)}\right)}\\
\derthreet{\eta_{l(i)}}{\log[f(\bsy_{(i)}|\bse)]}&=\left(-1\right)^{1-y_{l(i)}}\frac{\frac{\partial^2\phi(\eta^*_{l(i)})}{\partial\eta_{l(i)}^2}\Phi^2(\eta^*_{l(i)})-3\frac{\partial\phi(\eta^*_{l(i)})}{\partial\eta_{l(i)}}\Phi(\eta^*_{l(i)})\phi(\eta^*_{l(i)})+2\phi^3\left(\eta^*_{l(i)}\right)}{\Phi^3\left(\eta^*_{l(i)}\right)}\\
\derfourt{\eta_{l(i)}}{\log[f(\bsy_{(i)}|\bse)]}&=\frac{\frac{\partial^3\phi(\eta^*_{l(i)})}{\partial\eta_{l(i)}^3}\Phi^3\left(\eta^*_{l(i)}\right)-4\frac{\partial^2\phi(\eta^*_{l(i)})}{\partial\eta_{l(i)}^2}\Phi^2(\eta^*_{l(i)})\phi(\eta^*_{l(i)})}{\Phi^4\left(\eta^*_{l(i)}\right)}\\
&+\frac{12\frac{\partial\phi(\eta^*_{l(i)})}{\partial\eta_{l(i)}}\Phi(\eta^*_{l(i)})\phi^2(\eta^*_{l(i)})-3\left(\frac{\partial\phi(\eta^*_{l(i)})}{\partial\eta_{l(i)}}\right)^2\Phi^2(\eta^*_{l(i)})-6\phi^4(\eta^*_{l(i)})}{\Phi^4\left(\eta^*_{l(i)}\right)}
\end{align*}
where $\phi(\cdot)$ is the standard normal density function and we use the notational convention that
\begin{equation*}
\frac{\partial\phi(\eta^*_{l(i)})}{\partial\eta_{l(i)}}=\frac{\partial\phi(\eta_{l(i)})}{\partial\eta_{l(i)}}|_{\eta_{l(i)}=\eta^*_{l(i)}}
\end{equation*}
\subsection*{Normal Response}
For conditionally independent normal observations with $y_{l(i)}\mid\bse\sim N(\eta_{l(i)},\sigma^2_i)$,
\begin{align*}
\dert{\eta_{l(i)}}{\log[f(\bsy_{(i)}|\bse)]}&=\frac{y_{l(i)}-\eta_{l(i)}}{\sigma^2_i}\\
\dertwot{\eta_{l(i)}}{\log[f(\bsy_{(i)}|\bse)]}&=-\frac{1}{\sigma^2_i}\\
\derthreet{\eta_{l(i)}}{\log[f(\bsy_{(i)}|\bse)]}&= \derfourt{\eta_{l(i)}}{\log[f(\bsy_{(i)}|\bse)]} = 0
\end{align*}

More generally, let $\bsy_{(i)}\mid\bse\sim N_{n_{(i)}}(\boldsymbol\eta_{(i)},\bsR_{(i)})$, where $\boldsymbol\eta_{(i)}=\bsX_{(i)}\bsbeta_{(i)}+\bsZ_{(i)}\bse$, and hold the model parameters fixed. Writing $\ell_i=\log f(\bsy_{(i)}|\bse)$,
\[
\ell_i=-\frac{n_{(i)}}{2}\log(2\pi)-\frac12\log|\bsR_{(i)}|
-\frac12(\bsy_{(i)}-\boldsymbol\eta_{(i)})^{\prime}\bsR_{(i)}^{-1}(\bsy_{(i)}-\boldsymbol\eta_{(i)}).
\]
Componentwise differentiation gives
\[
\frac{\partial\ell_i}{\partial\eta_{l(i)}}=\sum_{m=1}^{n_{(i)}}(\bsR_{(i)}^{-1})_{lm}(y_{m(i)}-\eta_{m(i)}),
\qquad
\frac{\partial^2\ell_i}{\partial\eta_{l(i)}\partial\eta_{m(i)}}=-(\bsR_{(i)}^{-1})_{lm}.
\]
The same calculation in matrix form is
\begin{align*}
\nabla_{\boldsymbol\eta_{(i)}}\ell_i&=\bsR_{(i)}^{-1}(\bsy_{(i)}-\boldsymbol\eta_{(i)}),&
\nabla^2_{\boldsymbol\eta_{(i)}}\ell_i&=-\bsR_{(i)}^{-1},\\
\nabla_{\bse}\ell_i&=\bsZ_{(i)}^{\prime}\bsR_{(i)}^{-1}(\bsy_{(i)}-\boldsymbol\eta_{(i)}),&
-\nabla^2_{\bse}\ell_i&=\bsZ_{(i)}^{\prime}\bsR_{(i)}^{-1}\bsZ_{(i)}.
\end{align*}
In particular,
\[
-\sum_{l=1}^{n_{(i)}}\sum_{m=1}^{n_{(i)}}\frac{\partial^2\ell_i}{\partial\eta_{l(i)}\partial\eta_{m(i)}}\bsz_{l(i)}\bsz_{m(i)}^{\prime}
=\bsZ_{(i)}^{\prime}\bsR_{(i)}^{-1}\bsZ_{(i)}.
\]
All third and fourth predictor derivatives, including mixed derivatives, vanish for this normal block. The scalar formulas above are recovered when $\bsR_{(i)}=\sigma_i^2\boldsymbol I$, where $\boldsymbol I$ is the identity matrix.

 \section*{Appendix B: Calculation of Terms for the E step}\label{ssec:terms}
The calculations below apply to conditionally independent non-normal observations and to normal response blocks with any fixed residual covariance. The normal blocks have no third or fourth predictor derivatives, but their full curvature must be retained in $\bsSigma$ and $\bsSigma^{-1}$. A dependent non-normal response block would additionally require mixed third and fourth predictor derivatives. The E step calculations require the terms ${\partial}\bsSigma^{(c)}/{\partial c_k}$, ${\partial^2}\bsSigma^{(c)}/{\partial c_k\partial c_d}$, and $\partial H(\bse)/\partial \bse$. Furthermore, the calculation of the first two of these terms requires calculation of $\partial\widehat{\bse}^{(c)}/\partial{c_k}$ and ${\partial^2}\widehat{\bse}^{(c)}/{\partial c_k \partial c_d}$, both evaluated at $\bds{c}=0$. We first calculate $\partial\widehat{\bse}^{(c)}/\partial{c_k}$, following the method of \citet{riz09}. Let \[\kappa(\bse)=\left[\sum_{i=1}^q\log f\left(\bsy_{(i)}|\bse; \moparm\right)\right]+\log\left\{f\left(\bse\right)\right\}.\] Since, by definition, $\widehat{\bse}^{(c)} = \te{argmax}_{\bse}[\log\left\{f(\bsy_{(1)},\ldots,\bsy_{(q)},\bse)\right\} + \bds{c}^{\prime}H(\bse)]$, we have
\begin{align*}
\bds{0}&=\der{\bse}\left\{\kappa(\bse)+\bds{c}^{\prime}H(\bse)\right\}_{\bse=\widehat{\bse}^{(c)}}\\
&=\frac{\partial\kappa(\widehat{\bse}^{(c)})}{\partial\bse}+\frac{\partial \bds{c}^{\prime 	 }H(\widehat{\bse}^{(c)})}{\partial\bse}
\end{align*}
Taking the derivative with respect to $c_k$ yields
\begin{align*}
\frac{\partial^2\kappa(\widehat{\bse}^{(c)})}{\partial\bse\partial\bse^{\prime}}\frac{\partial\widehat{\bse}^{(c)}}{\partial c_k}+\frac{\partial\bep{k} H(\widehat{\bse}^{(c)})}{\partial\bse}+\left(\der{c_k}\left\{\frac{\partial H(\widehat{\bse}^{(c)})}{\partial\bse} \right\}  \right)^{\prime}\bds{c}=\bds{0}
\end{align*}
Solving for $\partial\widehat{\bse}^{(c)}/\partial{c_k}$ and evaluating at $c=0$ gives
\begin{align*}
\frac{\partial\widehat{\bse}^{(c)}}{\partial c_k}|_{\bds{c}=0}
=&\left(-\frac{\partial^2\kappa(\widehat{\bse})}{\partial\bse\partial\bse^{\prime}}\right)^{-1}\left(\frac{\partial \bep{k}H\left(\widehat{\bse}\right)}{\partial\bse}\right)\\
=&\bsSigma^{-1}\left(\frac{\partial \bep{k}H\left(\widehat{\bse}\right)}{\partial\bse}\right)
\end{align*}
We only need the terms ${\partial^2}\bsSigma^{(c)}/{\partial c_k\partial c_d}$ and ${\partial^2}\widehat{\bse}^{(c)}/{\partial c_k \partial c_d}$ for the case $H(\bse)=\bse$. These terms are used in the calculation of $\widetilde{\bsv}$. To find ${\partial^2}\widehat{\bse}^{(c)}/{\partial c_k \partial c_d}$ where $H(\bse)=\bse$, note
\begin{align*}
&\der{c_d}\left\{\frac{\partial^2\kappa(\widehat{\bse}^{(c)})}{\partial\bse\partial\bse^{\prime}}\frac{\partial\widehat{\bse}^{(c)}}{\partial c_k}+\frac{\partial \bep{k}H(\widehat{\bse}^{(c)})}{\partial\bse}+\left(\der{c_k}\left\{\frac{\partial H(\widehat{\bse}^{(c)})}{\partial\bse} \right\}  \right)^{\prime}\bds{c}\right\}=\bds{0}\\
&\Rightarrow\der{c_d}\left\{\frac{\partial^2\kappa(\widehat{\bse}^{(c)})}{\partial\bse\partial\bse^{\prime}}\frac{\partial\widehat{\bse}^{(c)}}{\partial c_k}+\be{k}\right\}=\bds{0}\\
&\Rightarrow\frac{\partial^2\kappa(\widehat{\bse}^{(c)})}{\partial\bse\partial\bse^{\prime}}\frac{\partial^2\widehat{\bse}^{(c)}}{\partial c_k\partial c_d}+\frac{\partial^3\kappa(\widehat{\bse}^{(c)})}{\partial\bse\partial\bse^{\prime}\partial\bse^{\prime}}\frac{\partial\widehat{\bse}^{(c)}}{\partial c_d}\frac{\partial\widehat{\bse}^{(c)}}{\partial c_k}=\bds{0}\\
&\stackrel{\bds{c}=\bds{0}}{\Rightarrow}\frac{\partial^2\widehat{\bse}^{(c)}}{\partial c_k\partial c_d}|_{\bds{c}=\bds{0}}=\left(\frac{\partial^2\kappa(\widehat{\bse})}{\partial\bse\partial\bse^{\prime}}\right)^{-1}\frac{\partial \bsSigma}{\partial \bse^{\prime}}\frac{\partial\widehat{\bse}}{\partial c_d}\frac{\partial\widehat{\bse}}{\partial c_k}\\
&\Rightarrow\frac{\partial^2\widehat{\bse}^{(c)}}{\partial c_k\partial c_d}|_{\bds{c}=\bds{0}}=\bsSigma^{-1}\left[\sum_{i=1}^q\sum_{l=1}^{n_{(i)}} \derthreet{\eta_{l(i)}}{\log[f(\bsy_{(i)}|\bse)]} \bsz_{l(i)}\bsz_{l(i)}^{\prime}\left(\bsz^{\prime}_{l(i)}\bsSigma^{-1}\be{d}\right)\right]\bsSigma^{-1}\be{k}.
\end{align*}
We may thus write
\begin{align*}
\frac{\partial\bsSigma^{(\bds{c})}}{\partial c_k}|_{\bds{c}=\bds{0}}
= -\sum_{i=1}^q\sum_{l=1}^{n_{(i)}}\left[\derthreet{\eta_{l(i)}}{\log[f(\bsy_{(i)}|\bse)]} \left(\bsz^{\prime}_{l(i)}\frac{\partial\widehat{\bse}^{(c)}}{\partial c_k}|_{\bds{c}=0}\right)\bsz_{l(i)}\bsz^{\prime}_{l(i)}\right]-\frac{\partial^2}{\partial\bse\partial\bse^{\prime}}\left[\bep{k}H(\widehat{\bse})\right]
\end{align*}
When $H(\bse)=\bse$, differentiate along $\bse=\widehat{\bse}^{(\bsc)}$ before evaluating at $\bsc=\bds{0}$:

\begin{align*}
\frac{\partial^2\bsSigma^{(\bds{c})}}{\partial c_k\partial c_d}=& \frac{\partial}{ \partial c_d}\left[-\sum_{i=1}^q\sum_{l=1}^{n_{(i)}}\derthreet{\eta_{l(i)}}{\log[f(\bsy_{(i)}|\bse)]} \left(\bsz^{\prime}_{l(i)}\frac{\partial\widehat{\bse}^{(c)}}{\partial c_k}\right)\bsz_{l(i)}\bsz^{\prime}_{l(i)}\right]_{\bds{c}=\bds{0}}\\
=&-\sum_{i=1}^q\sum_{l=1}^{n_{(i)}}\left\{\left[\bsz^{\prime}_{l(i)}\frac{\partial^2\widehat{\bse}^{(c)}}{\partial c_k\partial c_d}|_{\bds{c}=0}\derthreet{\eta_{l(i)}}{\log[f(\bsy_{(i)}|\bse)]}\right.\right.\\
&\left.\left.+\bsz^{\prime}_{l(i)}\left(\frac{\partial\widehat{\bse}^{(c)}}{\partial c_k}|_{\bds{c}=0}\right)\bsz^{\prime}_{l(i)}\left(\frac{\partial\widehat{\bse}^{(c)}}{\partial c_d}|_{\bds{c}=0}\right)\derfourt{\eta_{l(i)}}{\log[f(\bsy_{(i)}|\bse)]} \right]\bsz_{l(i)}\bsz^{\prime}_{l(i)}\right\}
\end{align*}

\subsection*{Equivalent matrix calculation}
For $H(\bse)=\bse$, write $\bsA=\bsSigma^{-1}$,
$\boldsymbol a_k=\bsA\be{k}$, and $\boldsymbol u_{ik}=\bsZ_{(i)}\boldsymbol a_k$.
Let $\boldsymbol d_{i,r}$ collect the scalar $r$th predictor derivatives of $\log f(\bsy_{(i)}|\bse)$ for $r=3,4$, evaluated at $\widehat{\bse}$; for a normal block these vectors are zero. All quantities below are evaluated at $\bsc=\bds{0}$. Using $\circ$ for componentwise multiplication and $\operatorname{diag}(\boldsymbol v)$ for the diagonal matrix with diagonal $\boldsymbol v$, the preceding calculations become
\begin{align*}
\left.\frac{\partial\widehat{\bse}^{(\bsc)}}{\partial c_k}\right|_{\bsc=\bds{0}}&=\boldsymbol a_k,\\
\boldsymbol a_{kd}:=\left.\frac{\partial^2\widehat{\bse}^{(\bsc)}}{\partial c_k\partial c_d}\right|_{\bsc=\bds{0}}
&=\bsA\sum_{i=1}^q\bsZ_{(i)}^{\prime}
 (\boldsymbol d_{i,3}\circ\boldsymbol u_{ik}\circ\boldsymbol u_{id}),\\
\bsSigma_k:=\left.\frac{\partial\bsSigma^{(\bsc)}}{\partial c_k}\right|_{\bsc=\bds{0}}
&=-\sum_{i=1}^q\bsZ_{(i)}^{\prime}
 \operatorname{diag}(\boldsymbol d_{i,3}\circ\boldsymbol u_{ik})\bsZ_{(i)},\\
\bsSigma_{kd}:=\left.\frac{\partial^2\bsSigma^{(\bsc)}}{\partial c_k\partial c_d}\right|_{\bsc=\bds{0}}
&=-\sum_{i=1}^q\bsZ_{(i)}^{\prime}\operatorname{diag}
 \left(\boldsymbol d_{i,4}\circ\boldsymbol u_{ik}\circ\boldsymbol u_{id}
 +\boldsymbol d_{i,3}\circ(\bsZ_{(i)}\boldsymbol a_{kd})\right)\bsZ_{(i)}.
\end{align*}
Thus (\ref{eq:etahat}) and (\ref{eq:varetahat}) can be evaluated as
\[
\widetilde b_k\approx\widehat b_k-\tfrac12\tr(\bsA\bsSigma_k),\qquad
\widetilde v_{kd}\approx A_{kd}-\tfrac12\tr(\bsA\bsSigma_{kd}-\bsA\bsSigma_d\bsA\bsSigma_k).
\]
Here $\widehat b_k$, $\widetilde b_k$, $\widetilde v_{kd}$ and $A_{kd}$ denote the corresponding vector and matrix entries. Although a correlated normal block contributes zero to $\boldsymbol d_{i,3}$ and $\boldsymbol d_{i,4}$, it still affects these corrections through the full matrix $\bsA$.

\subsection*{Additional derivatives for fixed-effect updates}
For the conditionally independent non-normal responses considered here, the fully exponential approximation for the integrals in the score function of $\bsbeta_{(i)}$ requires two additional terms: ${\partial H(\bse)}/{\partial\bse}$ and ${\partial^2 H(\bse)}/{\partial\bse\partial\bse^{\prime}}$. Note that these terms are not included in the summand over $l$. During the M step, the posterior mode, curvature and derivatives defining the posterior corrections remain fixed at the preceding E step parameter values and include all response blocks. The integrand $H$ and its derivatives are evaluated at the trial fixed-effect parameter values.
\begin{align*}
\frac{\partial}{\partial\bse}\left[\dert{\eta_{l(i)}}{\log[f(\bsy_{(i)}|\bse)]}  \right]&=\dertwot{\eta_{l(i)}}{\log[f(\bsy_{(i)}|\bse)]} \bsz_{l(i)}\\
\frac{\partial^2}{\partial\bse\partial\bse^{\prime}}\left[\dert{\eta_{l(i)}}{\log[f(\bsy_{(i)}|\bse)]}  \right]&=\derthreet{\eta_{l(i)}}{\log[f(\bsy_{(i)}|\bse)]}\bsz_{l(i)}\bsz^{\prime}_{l(i)}
\end{align*}
\section*{Acknowledgments}
This research was partially supported by the National Science Foundation under grant DRL-0909630, and by Arizona State University through a Dissertation Fellowship. Any opinions, findings, and conclusions or recommendations expressed in this material are those of the authors and do not reflect the views of the National Science Foundation, Adsurgo LLC, Arizona State University, or Westat. We would like to thank the anonymous reviewers, whose suggestions led to substantial improvements in the paper.

\end{document}